\documentclass[aip,reprint]{revtex4-1} 

\usepackage{amsmath}
\usepackage{amsfonts}   
\usepackage{amssymb}    
\usepackage{graphicx}   
\usepackage{xcolor}
\usepackage{bm}
\usepackage{secdot}		
\usepackage{mathptmx}
\usepackage{float}
\usepackage[utf8]{inputenc}
\usepackage{textcomp}
\usepackage[hang,flushmargin,bottom]{footmisc} 

\usepackage{booktabs}
\usepackage{setspace}
\usepackage{upgreek}
\usepackage[mathscr]{euscript}
\usepackage{tikz}
\usepackage{hyperref}

\newcommand{\revision}[1]{{#1}}

\newcommand{\pypx}[2]{\dfrac{\partial #1}{\partial #2}}

\newcommand{\pnypxn}[3]{\dfrac{\partial^{#3} #1}{\partial #2^{#3}}}
\newcommand{\vFvx}[2]{\dfrac{\updelta #1}{\updelta #2}}

\newcommand{\picst}[0]{\mathrm{\pi}}
\newcommand{\kBT}[0]{k_\mathrm{B}T}

\draft 

\begin{document}



\title{Nonclassical condensation pathways revealed by the multivariable theory of nucleation}



\author{Yijian Wu}
\email[]{yijian.wu@polytechnique.edu}
\affiliation{Laboratoire de Physique de la Matière Condensée, CNRS, Ecole polytechnique, Institut Polytechnique de Paris, 91120 Palaiseau, France}
\author{Thomas Philippe}
\email[]{thomas.philippe@polytechnique.edu}
\affiliation{Laboratoire de Physique de la Matière Condensée, CNRS, Ecole polytechnique, Institut Polytechnique de Paris, 91120 Palaiseau, France}

\date{\today}

\begin{abstract}
    Classical nucleation theory (CNT) provides a simple conceptual framework for nucleation but often fails to quantitatively reproduce experimental and numerical observations. In this work, we extend the theory by explicitly incorporating the multidimensional nature of nucleation and the coupled roles of kinetics and thermodynamics. Specifically, we treat the cluster density as an independent variable, within both sharp-interface and diffuse-interface descriptions. The kinetics are governed by dynamical density functional theory.
Applied to liquid condensation in the Lennard-Jones system, our self-consistent, parameter-free two-variable (size--density) and three-variable (size--interface width--density) models reveal nonclassical nucleation mechanisms. At low supersaturation, both models recover the classical picture, in which clusters nucleate and grow at the equilibrium liquid density. As supersaturation increases, nonclassical behavior emerges: the critical cluster density decreases, and the nucleation pathway involves concomitant evolution in cluster size, density, and, within the diffuse-interface description, interfacial width. 
Our diffuse-interface model reveals a rapid increase in interfacial diffuseness at high supersaturation. Near the spinodal limit, both models predict the critical cluster with diverging sizes, densities approaching that of the metastable initial phase, and vanishing work of formation, which provides a smooth connection between nucleation and spinodal decomposition. Comparison with molecular dynamics simulations demonstrates that both models substantially outperform CNT. However, the weak non-monotonic dependence of the critical cluster density observed at very low supersaturation is captured only by diffuse-interface models. Overall, our findings indicate that CNT should be applied only in the low-supersaturation regime, and our work provides a robust foundation for its refinement beyond this limit.

\end{abstract}


\maketitle 

\section{Introduction}
\label{sec:intro}
    A crucial feature of phase-separating systems is nucleation, wherein a new phase emerges through the formation and growth of a nucleus within a metastable mother phase. Nucleation is a rare event: the formation and growth of a cluster are driven by thermal noise until the critical cluster is reached, beyond which growth becomes deterministic and proceeds spontaneously. 

Classical nucleation theory (CNT)~\cite{volmer_kinetik_1939, becker_kinetische_1935, zeldovich_theory_1942, frenkel_kinetic_1946, Turnbull_1949, feder_homogeneous_1966} serves as the foundation for most nucleation models. CNT typically adopts the capillary model~\cite{Reguera_2003_revisited_problem, Reguera_2003}, where a nucleus is idealized as a spherical cluster separated from the mother phase by a sharp interface that incurs an energy penalty proportional to its interface area. The cluster size is the sole variable, while the density (or composition, in multicomponent systems) of the cluster is assumed to match the equilibrium value from the phase diagram. Despite the various refinements~\cite{Prestipino}, such as the Tolman correction, the capability of CNT to produce quantitative results remains debated. A common and significant shortcoming is that the nucleation rates predicted by CNT often deviate from experimental~\cite{iland_argon_2007,sinha_argon_2010} and simulation~\cite{Diemand_JCP2013,yoo_monte_2001} results by several orders of magnitude. One plausible reason for this discrepancy is the oversimplified representation of clusters: the one-variable, sharp-interface model in CNT may be insufficient to capture the nucleation process. 

To address this, diffuse-interface models~\cite{Cahn-Hilliard-1959, Oxtoby-Evans-1988, Granasy_1993, Ghosh-2011} have been proposed. These models offer a more realistic representation of nucleation by relaxing the sharp-interface approximation. In contrast to CNT, they allow for a smooth density transition between phases over a length scale of a few molecular diameters, known as the interface thickness. Moreover, the cluster density does not need to be the equilibrium density. A key advantage of diffuse-interface models is their ability to naturally account for interface curvature and thickness effects. This becomes particularly important at high supersaturation, where the critical cluster size approaches the interface thickness and CNT becomes unreliable. Lutsko applied such models to the condensation of Lennard-Jones particles using classical density functional theory (DFT) \cite{Lutsko_JCP2008} and the square-gradient approximation (SGA)~\cite{Lutsko-JCP2011}, showing fairly good agreement between theory and microscopic simulations. Similar conclusions were reached by Reguera \textit{et al.} in their DFT study of the argon fluid modeled with the Lennard-Jones potential~\cite{Reguera_2003}. However, diffuse-interface models introduce a significant challenge: the associated work of formation of a cluster becomes highly multidimensional because clusters are described by their full density profiles. This complexity makes the calculation of nucleation rate in this high-dimensional space extremely difficult. To date, such nucleation rate calculations have rarely been performed in practice. Notable recent attempts include work by Lutsko \textit{et al.} for a colloidal solution~\cite{Lutsko-JCP2024} and by Simeone \textit{et al.} within the SGA~\cite{Simeone_PRL2023}. While these approaches provide a promising theoretical foundation, their application to practical problems remains highly challenging.

Meanwhile, any advancement in nucleation theory needs confirmation from concrete observations. While homogeneous nucleation is notoriously difficult to probe experimentally, it can be systematically investigated by numerical simulations. In particular, recent advances in molecular dynamics (MD) simulations~\cite{Diemand_JCP2013, Angelil_JCP2014, Vega_Seeding_JCP2016, LAM_Ni3Al_2020, Lam_CV_JCP2023} allow the predictions of nucleation theories to be compared with microscopic modeling in terms of both nucleation rates and critical cluster properties. For instance, the authors of Refs.~\onlinecite{Diemand_JCP2013, Angelil_JCP2014} conducted large-scale brute-force MD simulations of Lennard-Jones condensation. They measured the steady-state nucleation rates over a range of supersaturations. They also found that the clusters could deviate from the spherical shape, and that their density could be significantly lower than the equilibrium liquid density from the phase diagram and increase with time as clusters develop. These findings provide clear evidence that the capillary approximation in CNT fails to capture the true nature of nucleation at moderate to high supersaturation. 

The objective of this paper is to build accessible and self-consistent multidimensional nucleation models. By self-consistent, we mean that all model parameters are physically determined without empirical fitting of nucleation simulation data; by accessible, we mean that the model can be straightforwardly applied to compute at least the critical cluster properties and nucleation rates. 
\revision{Throughout this work, we restrict our attention to homogeneous condensation, namely, the nucleation of a liquid droplet from a metastable vapor, even though the general multivariable formalism can be adapted to other first-order phase transitions by choosing suitable collective variables.}

Given the simplicity and intuitiveness of CNT, we first extend the classical theory and relax some of its approximations according to numerical observations. We retain the assumption of spherical cluster geometry. 
In our previous studies, to probe the critical cluster properties in low and intermediate supersaturation regimes, we conducted seeded and steering with aimless shooting MD simulations, and observed a decrease in critical cluster density with increasing supersaturation~\cite{Wu2026PRL}. The findings on critical cluster density motivate a key refinement of CNT: treating the cluster density as a free variable, rather than fixing it to the equilibrium value. Both the driving force for nucleation and the surface forming energy thereby depend on the cluster density. 
If we retain the capillary approximation, in which the cluster is separated from its mother phase by a sharp interface, then the nucleation model will have two variables: clusters are characterized by their size and density. Alternatively, if we relax the sharp-interface assumption, we can adopt a diffuse-interface radial density profile ansatz ~\cite{Lutsko_2008} with three variables: cluster density, a size-like variable, and an interface-width-related variable. 

The use of two variables to characterize the nucleus and account for less dense clusters is not new. It has been used by Schmelzer \textit{et al.} who named such an extension of CNT as the Generalized Gibbs theory~\cite{Schmelzer_2000,Schmelzer_2003,Schmelzer_2006,Schmelzer_2007,Schmelzer_2011,Baidakov_2000}. It was also proposed by Reguera \textit{et al.}~\cite{Reguera_2003_revisited_problem}, by Philippe \textit{et al.}~\cite{Philippe_JCP_2011,Philippe_2011_Phil_mag,Bonvalet_Phil_mag_2014}, by Lutsko \textit{et al.}~\cite{lutsko_two-parameter_2015}, and by Ghosh and Ghosh~\cite{Ghosh-2011}. It is sometimes referred to as the modified CNT. A priori, it was first introduced by Reiss and Shugard~\cite{Reiss_1976} in 1976 to reconcile the classical model with the Cahn--Hilliard theory of nucleation~\cite{Cahn-Hilliard-1959}, which is a diffuse-interface model. On the other hand, the three-variable model~\cite{Lutsko_2008} has rarely been investigated despite its apparent simplicity and potential advantages over the SGA formalism, as we shall see. In the present work, the two-variable and three-variable models are coupled to kinetics using recent developments in nucleation theories~\cite{alekseechkin_multivariable_2006,Lutsko-JCP2011} and compared with microscopic simulations.

Our workflow is as follows. First, in our extensions of the classical theory, the critical properties are obtained by locating the saddle point on the work of formation surface. Then, the nucleation rate and nucleation flux direction are determined by coupling the multivariable nucleation theory~\cite{langer_statistical_1969, alekseechkin_multivariable_2006} with the dynamical density functional theory, as in the mesoscopic nucleation model proposed by Lutsko~\cite{Lutsko-JCP2011}. Finally, predictions of our models are compared with atomistic simulations of Lennard-Jones condensation. An essential and brief overview of our theoretical and numerical work can be found in our previous publication~\cite{Wu2026PRL}, where we presented our innovative MD simulation strategy but only for the two-variable nucleation model with sharp interface. In the present paper, we will provide a more comprehensive and rigorous account of our theoretical work. Moreover, for a closer examination of the critical cluster properties, we perform new seeded MD simulations in the \textit{NVT} ensemble \cite{PHILIPPE2026108254} to collect a larger number of critical clusters in the intermediate regime of supersaturation. Section~\ref{sec:theory}, for the sake of generality, first presents the nucleation theory in an arbitrary multidimensional space and then derives its reduction to both two-variable and three-variable models. Section~\ref{sec:Results} discusses the predictions of our two-variable and three-variable models and compares them with MD simulation results and other theoretical predictions. In closing, a nonclassical picture of nucleation is discussed, focusing in particular on the behavior of the nucleation flux direction.

\section{Governing equations}
\label{sec:theory}
    \subsection{Multivariable nucleation theory}

Langer \cite{langer_statistical_1969} and Alekseechkin \cite{alekseechkin_multivariable_2006} independently developed multivariable nucleation theory in the phase space $\{\boldsymbol{x}\}$\revision{, where $\boldsymbol{x}$ denotes the collective variables or reaction coordinates chosen for the coarse-grained description of the nucleation process}. As we demonstrate (see Appendix~\ref{Appendix:Langer&Alekseechkin}), their formulas for the steady-state nucleation rate are almost equivalent, differing only in the determination of the parameter $f_0$ in the equilibrium density function of nuclei, 
\begin{equation}
    f_\mathrm{eq}\revision{(\boldsymbol{x})} = f_0\,\exp{\big(-\frac{\Delta\varOmega(\boldsymbol{x})}{k_\mathrm{B}T}\big)}.
\label{eq:EquilibriumDensityFunction}
\end{equation}
Here, $\Delta\varOmega$ represents the work of formation of a cluster, $k_\mathrm{B}$ is the Boltzmann constant, and $T$ is the temperature. 
In the following, we provide a brief review of multivariable nucleation theory \revision{for liquid condensation}, primarily using the notations introduced by Alekseechkin for simplicity in our subsequent discussions. $f_0$ is retained in the derivations, and its determination will be discussed later.

The Fokker--Planck equation for the system can be written as
\begin{equation}
    \pypx{f}{t} = -\boldsymbol{\nabla\cdot}\boldsymbol{J} , \quad 
    \boldsymbol{J} = -\mathbf{D}\boldsymbol{\nabla}f + \pypx{\boldsymbol{x}}{t}f,
\label{eq:FP_Alekseechkin}
\end{equation}
where $f=f(\boldsymbol{x},t)$ is the time-dependent density function of clusters in the $n$-dimensional phase space $\{\boldsymbol{x}\}$, $t$ is the time, $\boldsymbol{J}$ is the nucleation flux vector in the phase space, and $\mathbf{D}$ is the diffusivity matrix. We emphasize that $\mathbf{D}$ is a matrix corresponding to the phase space, It is not, but related to, the diffusion coefficient in the real space (see Subsection~\ref{DDFT}).

The phase space is divided into a \revision{subcritical side (one subcritical liquid cluster in the vapor)} and a \revision{postcritical side (one postcritical liquid cluster in the vapor)}. We choose the unit for the nucleation rate as per volume per time, so $f$ is normalized in the \revision{subcritical (subC)} side as
\begin{equation}
    \int_\mathrm{subC} f \mathrm{d}\boldsymbol{x} = \rho_\mathrm{cluster}, 
\label{eq:f0_normalization}
\end{equation}
with $\rho_\mathrm{cluster}$ the number density of subcritical clusters. In this sense, $f/\rho_\mathrm{cluster}$ is the probability density function of subcritical clusters. However, $\rho_\mathrm{cluster}$ is unknown. 

At equilibrium, the nucleation flux vanishes, and by considering Eq.~(\ref{eq:EquilibriumDensityFunction}), it follows that
\begin{equation}
    \pypx{\boldsymbol{x}}{t} = -\frac{1}{\kBT}\mathbf{D}\boldsymbol{\nabla}\Delta\varOmega.
\label{eq:dotx_Alekseechkin}
\end{equation}
The nucleation trajectory is supposed to pass through the saddle point of the work of formation surface, $(\boldsymbol{x}_\mathrm{c}, \Delta\varOmega_\mathrm{c})$. In its vicinity, the work of formation can be approximated as 
\begin{equation}
    \Delta\varOmega = \Delta\varOmega_\mathrm{c}+\frac{1}{2}(\boldsymbol{x}-\boldsymbol{x}_\mathrm{c})^\mathrm{T}\mathbf{H}(\boldsymbol{x}-\boldsymbol{x}_\mathrm{c}),
\label{eq:expansion WOF}
\end{equation}
with $\mathbf{H}$ the Hessian matrix of the work of formation at the saddle point. Substituting this into Eq.~(\ref{eq:dotx_Alekseechkin}), we get 
\begin{equation}
    \pypx{\boldsymbol{x}}{t} = -\frac{1}{\kBT}\mathbf{DH}(\boldsymbol{x}-\boldsymbol{x}_\mathrm{c}) \equiv -\frac{1}{\kBT}\mathbf{Z}(\boldsymbol{x}-\boldsymbol{x}_\mathrm{c}).
\label{eq:dotxDH_Alekseechkin}
\end{equation}
Here, the matrix $\mathbf{Z} \equiv \mathbf{DH}$ is defined in the same way as Alekseechkin~\cite{alekseechkin_multivariable_2006}.  

The general expression for the steady-state nucleation rate is given by \cite{alekseechkin_multivariable_2006,langer_statistical_1969}
\begin{equation}
    I=f_0 \exp{\big(-\dfrac{\Delta\varOmega_\mathrm{c}}{k_\mathrm{B}T}\big)} \big(2\picst k_\mathrm{B}T\big)^{\frac{n-2}{2}} \frac{|{\lambda_\mathbf{Z}}_1|}{\sqrt{|\det \mathbf{H}|}}.
\label{eq:I}
\end{equation}
This formula holds as long as the determinant of $\mathbf{H}$ is non-zero. Here, ${\lambda_\mathbf{Z}}_1$ is the only negative eigenvalue of matrix $\mathbf{Z}$, and its corresponding eigenvector ${\boldsymbol{v}_\mathbf{Z}}_1$ points along the unstable direction of nucleation, which is indeed the nucleation flux direction. $|{\lambda_\mathbf{Z}}_1| / (k_\mathrm{B}T)$ can be regarded as the exponential growth rate of the system along ${\boldsymbol{v}_\mathbf{Z}}_1$. The nucleation current therefore results from a complex interplay between thermodynamics ($\mathbf{H}$) and kinetics ($\mathbf{D}$) that is encoded in the $\mathbf{Z}$ matrix.

In closing, for the distribution density parameter $f_0$, although Alekseechkin~\cite{alekseechkin_multivariable_2006} and Langer~\cite{langer_statistical_1969} both proposed analytical expressions, their assumptions do not always hold in our studies (see Appendix~\ref{Appendix:f0}). We propose to employ mass conservation to calculate $f_0$,
\begin{equation}
    \int_\mathrm{subC} f_\mathrm{eq}(\boldsymbol{x}) \,q(\boldsymbol{x}) \,\mathrm{d}\boldsymbol{x} = \rho_0,
\label{eq:f0_MassConservation}
\end{equation}
where $q(\boldsymbol{x})$ is the number of monomers in a cluster, and $\rho_0$ is the initial density of the system. Therefore,
\begin{equation}
    f_0 = \dfrac{\rho_0}{\displaystyle\int_\mathrm{subC} \exp{\big(-\frac{\Delta\varOmega(\boldsymbol{x})}{k_\mathrm{B}T}\big) \,q(\boldsymbol{x}) \,\mathrm{d}\boldsymbol{x}}}.
\label{eq:f0}
\end{equation}

\subsection{Coupling to dynamical density functional theory}    
\label{DDFT}
In principle, the diffusivity matrix $\mathbf{D}$ of the Fokker--Planck equation can be derived from either microscopic considerations, such as the attachment and detachment rates of monomers in the cluster growth process~\cite{kalikmanov_nucleation_2013}, or macroscopic equations \cite{alekseechkin_multivariable_2006}. It is the latter method that we use in the following. In its pioneering description of classical fluids~\cite{Evans01041979}, Evans assumed that in the real space $\{\mathbf{r}\}$ the driving force for particle diffusion is related to the gradient of the chemical potential,
\begin{equation}
\label{eq:diff_flux}
    \boldsymbol{j} (\mathbf{r}) = -\frac{D_\mathrm{diff}}{\kBT}\widetilde{\rho}(\mathbf{r}) \boldsymbol{\nabla} \mu(\mathbf{r}),
\end{equation}
where $D_\mathrm{diff}$ is the diffusion constant and $\mu(\mathbf{r})$ is the chemical potential calculated as the functional derivative of the total Helmholtz free energy of the system with respect to the density field $\widetilde{\rho}(\mathbf{r})$, 
\begin{equation*}
    \mu(\mathbf{r}) = \vFvx{F[\widetilde{\rho}]}{\widetilde{\rho}(\mathbf{r})} .
\end{equation*}
Together with Eq.~(\ref{eq:diff_flux}), the continuity equation
\begin{equation}
    \pypx{\widetilde{\rho}(\mathbf{r}\revision{,t})}{t} = - \boldsymbol{\nabla\cdot} \boldsymbol{j}
\end{equation}
provides a basis for the formulation of the dynamical density functional theory (DDFT). DDFT was then formally derived by Marconi and Tarazona~\cite{Marconi_2000}, and Archer and Evans~\cite{Archer-Evans-DDFT-2004}. It asserts that the most likely nucleation pathway connecting the metastable and stable phases passes through the saddle point of the free energy landscape~\cite{lutsko_communication_2011}. Within DDFT, the evolution of the time-dependent density profile $\widetilde{\rho}(\mathbf{r})$ is governed by 
\begin{equation}
    \pypx{\widetilde{\rho}(\mathbf{r}\revision{,t})}{t} = \frac{D_\mathrm{diff}}{\kBT}\boldsymbol{\nabla\cdot}\Bigg(\widetilde{\rho}(\mathbf{r}\revision{,t})\boldsymbol{\nabla} \vFvx{F[\widetilde{\rho}]}{\widetilde{\rho}(\mathbf{r}\revision{,t})}\Bigg) .
\label{eq:DDFT}
\end{equation}

Following closely the method proposed by Lutsko~\cite{lutsko_communication_2011}, we parameterize the density profile to derive a dynamical equation in the phase space $\{\boldsymbol{x}\}$ for our nucleation theory. We define $n(r,t)$ as the time-dependent integrated density within a sphere of radius $r$,
\begin{equation*}
    n(r,t) = \int_{r'<r} \widetilde{\rho}(\mathbf{r'},t)\, \mathrm{d}\mathbf{r'} .
\end{equation*}
In homogeneous nucleation, due to the spherical symmetry, this becomes 
\begin{equation}
    n(r,t) = \int_{0}^{r} 4\picst {r'}^2\widetilde{\rho}(r',t)\, \mathrm{d}r' ,
\label{eq:n_DDFT}
\end{equation}
and $\widetilde{\rho}(\mathbf{r})$ and $\mu(\mathbf{r})$ depend only on the radial coordinate $r$, simplified to $\widetilde{\rho}(r)$ and $\mu(r)$, respectively. 
By integrating Eq.~(\ref{eq:DDFT}) as Eq.~(\ref{eq:n_DDFT}) and applying Gauss' theorem, we obtain
\begin{equation}
    \dfrac{1}{4\picst r^2\widetilde{\rho}(r\revision{,t})} \pypx{n(r\revision{,t})}{t} = \dfrac{D_\mathrm{diff}}{\kBT}\pypx{\mu(r)}{r} .
\label{eq:DDFT1}
\end{equation}

The density profile is then parameterized as $\widetilde{\rho}(\revision{r},t)\longmapsto\widetilde{\rho}(\revision{r},\boldsymbol{x}(t))$, with $\boldsymbol{x}$ the variables in the phase space. This parameterization is valid since $\boldsymbol{x}$ is capable of fully describing the density profile under the assumed model. Correspondingly, the particle number $n(r,t)$ becomes $n(r,\boldsymbol{x}(t))$ and 
\begin{equation}
    \pypx{n}{t} = (\pypx{n}{\boldsymbol{x}})^\mathrm{T}\pypx{\boldsymbol{x}}{t} .
\label{eq:pnpt}
\end{equation}
Substituting Eq.~(\ref{eq:pnpt}) into Eq.~(\ref{eq:DDFT1}), multiplying from the left by $\pypx{n}{\boldsymbol{x}}$, and integrating with respect to $r$ from $0$ to infinity, we obtain
\begin{equation}
    \mathbf{g}\pypx{\boldsymbol{x}}{t} = \dfrac{D_\mathrm{diff}}{\kBT} \int_0^\infty \pypx{\mu(r)}{r} \pypx{n}{\boldsymbol{x}} \,\mathrm{d}r ,
\label{eq:DDFT2}
\end{equation}
where the elements of the matrix $\mathbf{g}$ are given by 
\begin{equation}
    \mathrm{g}_{ij} = \int_0^\infty \dfrac{1}{4\picst r^2\widetilde{\rho}(r)} \pypx{n}{x_i}  \pypx{n}{x_j} \,\mathrm{d}r .
\label{eq:gij_DDFT}
\end{equation}
This is the $\mathbf{g}$ matrix provided by Lutsko~\cite{lutsko_communication_2011} in its so-called mesoscopic theory of nucleation. It can be easily connected to Alekseechkin's theory. The right-hand side of Eq.~(\ref{eq:DDFT2}) can be evaluated using integration by parts, yielding
\begin{equation*}
\begin{aligned}
    \int_0^\infty \pypx{\mu(r)}{r} \pypx{n}{\boldsymbol{x}} \,\mathrm{d}r &= 
    \pypx{n}{\boldsymbol{x}}\mu(r)\bigg|_{r=\infty} - \pypx{n}{\boldsymbol{x}}\mu(r)\bigg|_{r=0} \\
    &\; - \int_0^\infty \mu(r) \frac{\partial ^2 n}{\partial \boldsymbol{x} \partial r} \,\mathrm{d}r \\ 
    &= \pypx{n}{\boldsymbol{x}}\mu(r)\bigg|_{r=\infty} - \pypx{n}{\boldsymbol{x}}\mu(r)\bigg|_{r=0} \\
    &\; - \int_0^\infty \vFvx{F[\widetilde{\rho}]}{\widetilde{\rho}(r)} \pypx{\widetilde{\rho}(r)}{\boldsymbol{x}} \,\mathrm{d}r .
\end{aligned} 
\end{equation*}
At $r=\infty$, the system reaches the homogeneous initial state with chemical potential $\mu_\mathrm{i}$, and its total particle number is $n_\mathrm{total}$. Therefore, the first term of the above equation simplifies to $\frac{\partial}{\partial \boldsymbol{x}} (n_\mathrm{total}\mu_\mathrm{i})$. The second term vanishes because at $r=0$, $\frac{\partial n}{\partial \boldsymbol{x}} = \frac{4}{3}\picst r^3 \frac{\partial \widetilde{\rho}(r)}{\partial \boldsymbol{x}}$, and $\frac{\partial \widetilde{\rho}(r)}{\partial \boldsymbol{x}}$ remains finite. The third term can be handled using the chain rule of derivative, yielding $\frac{\partial F}{\partial \boldsymbol{x}}$. 
Thus, we have 
\begin{equation*}
    \int_0^\infty \pypx{\mu(r)}{r} \pypx{n}{\boldsymbol{x}} \,\mathrm{d}r = \pypx{(n_\mathrm{total}\mu_\mathrm{i})}{\boldsymbol{x}} - \pypx{F}{\boldsymbol{x}} = -\pypx{\Delta\varOmega}{\boldsymbol{x}} ,
\end{equation*}
where $\Delta\varOmega = F - n_\mathrm{total}\mu_\mathrm{i}$ represents the work of formation. Substituting this result into Eq.~(\ref{eq:DDFT2}), we get
\begin{equation}
    \mathbf{g}\pypx{\boldsymbol{x}}{t} = -\dfrac{D_\mathrm{diff}}{\kBT}\pypx{\Delta\varOmega}{\boldsymbol{x}} = -\dfrac{D_\mathrm{diff}}{\kBT}\boldsymbol{\nabla}\Delta\varOmega .
\label{eq:DDFT3}
\end{equation}
Comparing this with Eq.~(\ref{eq:dotx_Alekseechkin}), we can directly identify the diffusivity matrix as $\mathbf{D}=D_\mathrm{diff}\,\mathbf{g^\mathrm{-1}}$. Therefore, the matrix $\mathbf{Z}$ is given by
\begin{equation}
    \mathbf{Z}=D_\mathrm{diff}\,\mathbf{g^\mathrm{-1}}\mathbf{H}, 
\label{eq:DDFT_Z}
\end{equation}
which couples the nucleation theory to the dynamics imposed by DDFT. 

\subsection{Two-variable and three-variable cases}
\label{2-var and 3-var}
We simplify the application of the multivariable nucleation theory for practice, as two-variable and three-variable models.

In our two-variable model, we adopt the capillary approximation from CNT and describe the nucleation properties using two key variables: the cluster is assumed to be spherical with a radius $R$, and its internal number density $\rho$ is considered homogeneous. The surrounding medium is also homogeneous with the initial density $\rho_0$. Thus, the phase-space variables become $\boldsymbol{x}=(R,\rho)^\mathrm{T}$.

In this framework, the monomer number function is given by $q = 4\picst \rho R^3/3$. The work of formation, $\Delta\varOmega$, consisting of contributions from the energetically favorable bulk free energy difference between the two phases and the unfavorable surface forming energy, is given by
\begin{equation}
    \Delta\varOmega=-\frac{4}{3}\mathrm{\pi}{R}^3g_n+4\mathrm{\pi}{R}^2\gamma.
\label{eq:WorkOfFormation_2var}
\end{equation}
Here, $g_n$ is the driving force for nucleation, classically defined as \cite{Cahn-Hilliard-1959}
\begin{equation}
    g_n = \mathcal{F}(\rho_0)+\pypx{\mathcal{F}}{\rho}\Bigg|_{\rho=\rho_0}(\rho-\rho_0)-\mathcal{F}(\rho),
\label{eq:gn}
\end{equation}
where $\mathcal{F}$ is the Helmholtz free energy per volume. The surface tension of the gas--liquid interface, $\gamma$, is regarded as density-dependent and inspired by the SGA,
\begin{equation}
    \gamma = \kappa(\rho-\rho_0)^2.
\label{eq:kappa}
\end{equation}
The factor $\kappa$ is supposed to be only temperature-dependent. It can be calculated from the surface tension at equilibrium~\cite{Schmelzer_2007,Philippe_2011_Phil_mag,Ghosh-2011,lutsko_two-parameter_2015,Philippe_2024} using 
\begin{equation*}
    \kappa = \frac{\gamma^\mathrm{eq}}{(\rho^\mathrm{eq}_1-\rho^\mathrm{eq}_2)^2} ,
\end{equation*}
where $\rho^\mathrm{eq}_1$ and $\rho^\mathrm{eq}_2$ are the phase diagram equilibrium gas and liquid densities, respectively. In this formulation, we neglect curvature corrections such as the Tolman length~\cite{Tolman_1949,Angelil_JCP2014,Dillmann_1991,Prestipino}, as the Tolman length should remain small in this model~\cite{Wu2026PRL}. 
It is important to note that, in contrast to CNT, where the density is fixed as the equilibrium density, the two-variable nucleation theory treats both the driving force, $g_n$, and surface tension, $\gamma$, as functions of the cluster density, $\rho$.

The critical cluster corresponds to the saddle point of the work of formation, denoted as $(R_\mathrm{c},\,\rho_\mathrm{c})$. Using the assumptions mentioned earlier, the critical cluster density, $\rho_\mathrm{c}$, satisfies the following implicit relation: 
\begin{equation*}
    3g_n(\rho_\mathrm{c}) = \pypx{g_n}{\rho}\Bigg|_{\rho=\rho_\mathrm{c}}(\rho_\mathrm{c}-\rho_0).
\end{equation*}
This equation implies that $\rho_\mathrm{c}$ is independent of the surface tension coefficient, $\kappa$.

The Hessian matrix of the work of formation is given by
\begin{equation*}
    \mathbf{H} = 
    \begin{pmatrix}
        \mathrm{H}_{RR} & \mathrm{H}_{R\rho}  \\[7pt]
        \mathrm{H}_{\rho R} & \mathrm{H}_{\rho\rho}
    \end{pmatrix}
    ,
\end{equation*}
with 
\begin{equation*}
    \begin{aligned}
        \mathrm{H}_{RR} &= \pnypxn{\Delta\varOmega}{R}{2} = -8\picst Rg_n+8\picst \kappa(\rho-\rho_0)^2,  \\
        \mathrm{H}_{\rho \rho} &= \pnypxn{\Delta\varOmega}{\rho}{2} = -\frac{4}{3}\picst R^3 \pnypxn{g_n}{\rho}{2} + 8\picst R^2\kappa,  \\
        \mathrm{H}_{R\rho} = \mathrm{H}_{\rho R} &= \dfrac{\partial ^2 \Delta\varOmega}{\partial \rho \partial R} = -4\picst R^2\pypx{g_n}{\rho} + 16\picst R\kappa(\rho-\rho_0).
    \end{aligned}
\end{equation*}
The matrix $\mathbf{g}$ is given by 
\begin{equation*}
    \mathbf{g} = 
    \begin{pmatrix}
        \mathrm{g}_{RR} & \mathrm{g}_{R\rho}  \\[7pt]
        \mathrm{g}_{\rho R} & \mathrm{g}_{\rho\rho}
    \end{pmatrix}
    ,
\end{equation*}
with \cite{lutsko_two-parameter_2015}
\begin{equation*}
    \begin{aligned}
        \mathrm{g}_{RR} &= 4\picst R^3 \dfrac{(\rho-\rho_0)^2}{\rho_0}  ,\\
        \mathrm{g}_{\rho \rho} &= \frac{4}{9}\picst R^5 \big(\frac{1}{5\rho}+\frac{1}{\rho_0}\big)  ,\\
        \mathrm{g}_{R\rho} = \mathrm{g}_{\rho R} &= 4\picst R^4 \, \dfrac{\rho-\rho_0}{3\rho_0}. 
    \end{aligned}
\end{equation*}

In our three-variable model, we relax the capillary approximation and adopt the modified sigmoid function \cite{Lutsko_2008} to describe the diffuse-interface density profile $\widetilde{\rho} (r)$,
\begin{equation}
    \widetilde{\rho}(r)=\rho_0+(\rho-\rho_0)\frac{1+br}{1+(br)^2}\frac{1-\tanh((r-R)/w)}{1-\tanh(-R/w)},
\label{eq:ModifiedSigmoid}
\end{equation}
with $b=2/(w(\exp{(2R/w)}+1))$. $\rho$ is the density at the center of the cluster, $R$ is cluster-size-related, and $w$ represents the liquid/gas interface width. The phase-space variables then become $\boldsymbol{x}=(R,w,\rho)^\mathrm{T}$.  It should be noted that in diffuse-interface models, usually people define the radius of the cluster as where the density is equal to the average of the densities at the center of the cluster and very far from the cluster, i.e., $(\rho+\rho_0)/2$. This radius, denoted as $\bar{R}$, is not the same as the space variable $R$. Therefore, the monomer number function is given by $q = 4\picst \displaystyle\int_0^{\bar{R}} r^2 \widetilde{\rho}(r) \mathrm{d}r$. 

Combined with the SGA \cite{Cahn-Hilliard-1959,Philippe_JCP_2011,Lutsko-JCP2011}, the work of formation is given by the functional
\begin{equation}
    \Delta \varOmega = 4\mathrm{\pi} \int_0^\infty \bigg[-g_n\left( \widetilde{\rho} \left( r \right)  \right) + \frac{K}{2} \big( \frac{\partial \widetilde{\rho}(r)}{\partial r}  \big)^2 \bigg] r^2 \,\mathrm{d}r,
\label{eq:WorkOfFormation_3var}
\end{equation}
with $K$ determined from the surface energy of the planar interface at equilibrium, as to calibrate phase-field models \cite{Luneville,Luneville2024,TISSOT2023119303,Simeone_PRL2023}. The critical cluster corresponds as well to the saddle point of the work of formation, $(R_\mathrm{c}, w_\mathrm{c}, \rho_\mathrm{c})$. The calculation of the saddle point and matrices $\mathbf{H}$ and $\mathbf{g}$ is carried out numerically.

\section{Results and discussion}
\label{sec:Results}
We examine our models by investigating systems governed by the well-known 12-6 Lennard-Jones potential,
\begin{equation*}
    v_\mathrm{LJ}(r) = 4\epsilon\Big(\big(\dfrac{\sigma}{r}\big)^{12} - \big(\dfrac{\sigma}{r}\big)^6\Big),
\end{equation*}
where $\epsilon$ and $\sigma$ represent the energy and length parameters, respectively. The potential reaches its minimum at $r_\mathrm{min}=2^{1/6}\sigma$ with $v_\mathrm{LJ}(r_\mathrm{min})=-\epsilon$. The long-ranged attractive component of the potential ensures the phase separation \cite{hansen_theory_2013}.

In this work, we work at $T=0.8\epsilon/k_\mathrm{B}$. The equilibrium surface tension \cite{baidakov_metastable_2007} is $\gamma^\mathrm{eq} = 0.8633\epsilon/\sigma^{2}$.  The Helmholtz free energy of the homogeneous phase of the Lennard-Jones system is calculated using perturbation theory \cite{chandler_equilibrium_1970,weeks_role_1971,andersen_relationship_1971,ree_equilibrium_1976,kang_perturbation_1985} within classical density functional theory (DFT), following Ref.~\onlinecite{lutsko_effect_2005}. Details of the free energy calculation within DFT are provided in Appendix \ref{Appendix:Recipe F}. It is expected to give an accurate description of the driving force for nucleation, which is a key quantity of any nucleation model. Another approach would be to use equations of state (EOS) to describe the free energy and the other thermodynamic properties. There are many different EOS for the Lennard-Jones system, and a detailed review can be found in Ref.~\onlinecite{stephan_review_2020}. The most widely used EOS are the ones of Kolafa and Nezbeda \cite{kolafa_lennard-jones_1994} and the ones of Johnson, Zollweg, and Gubbins (JZG) \cite{johnson_lennard-jones_1993}. EOS are known for producing phase diagrams that closely match simulation results, and they offer a simpler approach for calculating free energy compared to DFT. However, these two EOS are both empirical. While their functional forms are informed by physical principles, the parameters within these equations are fitted to match the phase diagrams from simulation results. In this work, we adopt DFT for calculating free energy since it is a more physically derived model. Moreover, its predicted equilibrium densities ($\rho^\mathrm{eq}_1 = 0.0063\sigma^{-3}$, $\rho^\mathrm{eq}_2 = 0.7847\sigma^{-3}$) are in very good agreement with microscopic simulations at $T=0.8\epsilon/k_\mathrm{B}$. Our previous studies \cite{PHILIPPE2026108254} also indicate that, at this temperature, DFT and these two EOS provide very similar and accurate predictions, compared with seeded MD simulations, on critical cluster size. 

In Ref.~\onlinecite{Wu2026PRL}, a series of \textit{NVT}-to-\textit{NPT} simulations were performed to measure the cluster growth rate and then extract the diffusion coefficient from Eqs.~(\ref{eq:dotxDH_Alekseechkin}) and (\ref{eq:DDFT_Z}). In practice, the \textit{NVT} seeding technique was first employed to stabilize liquid clusters in equilibrium with the surrounding gas. Once a cluster reached equilibrium, the simulation was switched to the \textit{NPT} ensemble to emulate an infinite system. In this stage, the pressure was fixed at the time-averaged virial pressure obtained from the preceding \textit{NVT} simulation, while the system volume was allowed to fluctuate to maintain the prescribed pressure. However, because pressure fluctuations are relatively large in Lennard-Jones systems of such sizes, the capability of \textit{NPT} simulations to faithfully emulate an infinite system is open to question. In the present work, we therefore employ another strategy to measure the diffusion coefficient. A postcritical liquid cluster of radius $R$ is initially inserted into a gas of density $\rho_0$, and then an \textit{NVT} simulation is performed to simulate the cluster growth. For sufficiently large clusters, growth can be modeled using only the cluster radius as a parameter, as in the traditional picture. Thus, the growth equation reads:
\begin{equation}
\label{growth_1D}
    \frac{\mathrm{d}R}{\mathrm{d}t} = \frac{2\gamma^{\mathrm{eq}} \rho_0 D_\mathrm{diff}}{k_\mathrm{B}T(\rho_2^{\mathrm{eq}})^2 R }\left( \frac{1}{R_\mathrm{c}^\mathrm{CNT}} - \frac{1}{R}\right).
\end{equation}

This classical result~\cite{Philippe_JCP2016} can also be derived within the DDFT framework by invoking the one-variable description~\cite{lutsko_dynamical_2012}, which proves the internal consistency of the approach. Here, $R_\mathrm{c}^\mathrm{CNT}$ denotes the critical cluster radius calculated by CNT at the gas density $\rho_0$, and it is nearly constant during the simulated growth regime in a sufficiently large system. At $T=0.8\epsilon/k_\mathrm{B}$, a direct fit of Eq.~(\ref{growth_1D}) to MD results yields a diffusion constant of $D_\mathrm{diff} = 0.237\sigma\sqrt{\epsilon/m}$, which is one order of magnitude larger than the value reported in Ref.~\onlinecite{Wu2026PRL}.

We calculate the nucleation properties for condensation processes with initial density $\rho_0$ varying from the equilibrium gas density, $\rho^\mathrm{eq}_1$, to the spinodal density, $\rho^\mathrm{s}_1$. All properties are in the Lennard-Jones units, namely $\epsilon$, $\sigma$, $k_\mathrm{B}$, and the monomer mass, $m$. A detailed derivation of the dimensionless nucleation rate and other key properties is provided in Appendix~\ref{Appendix:Dimensionless}. 

For comparison purposes, the predictions of our models will be compared with those from CNT and the full SGA model, as well as with MD simulation results reported in Refs.~\onlinecite{Diemand_JCP2013,Wu2026PRL} and obtained from new simulations. In the full SGA model, the work of formation is the same functional as the one in our three-variable model, while its density profile has no ansatz. The critical density profile corresponds to the saddle point of the work of formation and needs to be determined by solving the Euler--Lagrange equation. 

\revision{In the two-variable model, the Tolman correction~\cite{Tolman_1949}, which accounts for the curvature dependence of surface tension, will be neglected. Our previous study suggested that this correction should remain small for Lennard-Jones liquid~\cite{Wu2026PRL}.  In contrast, this curvature-induced effect is implicitly considered in both the three-variable and full SGA diffuse-interface models.}

\subsection{Critical cluster properties}
Predictions on critical cluster radius, interfacial width, density, and work of formation from our two-variable and three-variable models, CNT, and the full SGA model are reported in Fig.~\ref{fig:R_c,rho_c,DeltaOmega_c}. 

\begin{figure}[thb]
\begin{tikzpicture}
\node (fig1) at (0,0) {\includegraphics[trim={0.3cm 0cm 0.3cm 0cm},clip,width=0.45\textwidth]{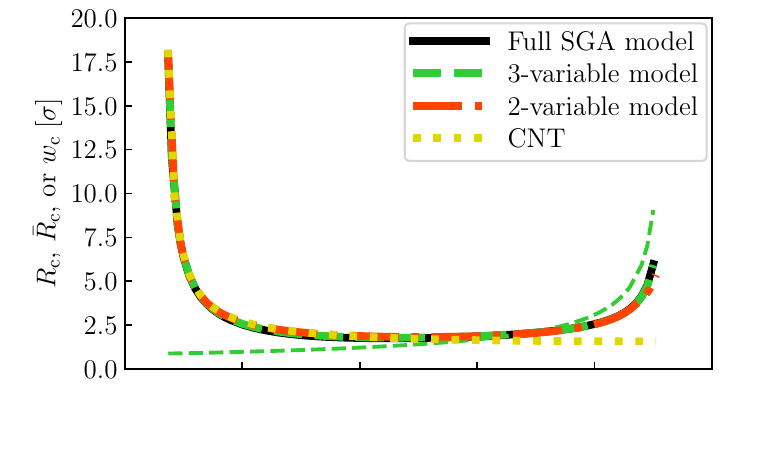}};
\node[] at (-4,2.5) {(a)};
\node (fig2) at (0,-4.13) {\includegraphics[trim={0.3cm 0cm 0.3cm 0cm},clip,width=0.45\textwidth]{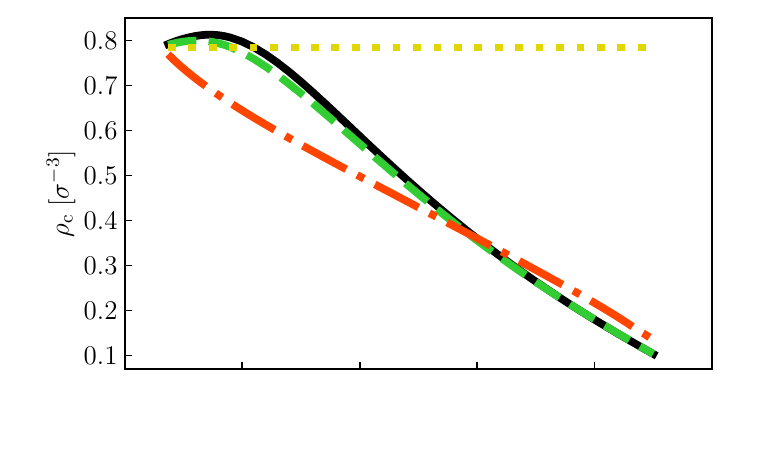}};
\node[] at (-4,-1.8) {(b)};
\node (fig3) at (0,-8.2) {\includegraphics[trim={0.3cm 0cm 0.3cm 0cm},clip,width=0.45\textwidth]{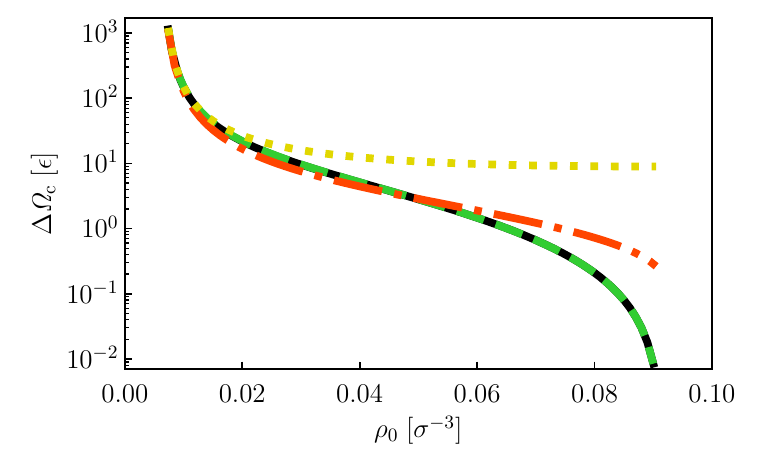}};
\node[] at (-4,-6) {(c)};
\end{tikzpicture}
\caption{Evolutions of the (a) radius and interfacial width, (b) density, and (c) work of formation of critical cluster as functions of the initial density at $T=0.8\epsilon/k_\mathrm{B}$. Predictions from our two-variable model, three-variable model, CNT, and the full SGA model are represented by red dash-dotted lines, green dashed lines, yellow dotted lines, and black solid lines, respectively. The thinner green dashed line in panel (a) represents the interfacial width of critical cluster predicted by the three-variable model. }
\label{fig:R_c,rho_c,DeltaOmega_c}
\end{figure}
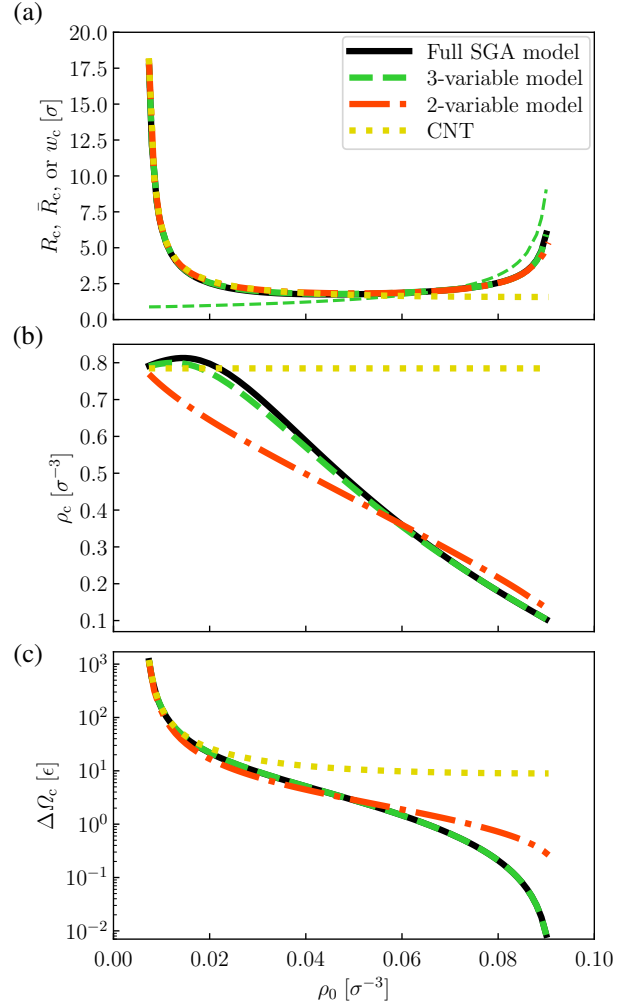

In the traditional application of CNT, the critical cluster density is assumed to be equal to the equilibrium liquid density, 
\begin{equation*}
    \rho_\mathrm{c}^\mathrm{CNT} = \rho_2^\mathrm{eq}.
\end{equation*}
Therefore, the driving force of nucleation is given by $g_n^\mathrm{CNT}=g_n(\rho^\mathrm{eq}_2)$. It should be noted that people usually use the chemical potential difference under the supersaturated pressure to describe the nucleation driving force. Our expression is indeed equivalent to it (see Appendix~\ref{Appendix:Delta mu in CNT}).
The surface tension is assumed to be $\gamma^\mathrm{eq}$. Consequently, the radius and the work of formation for the critical cluster are given by
\begin{equation*}
    R_\mathrm{c}^\mathrm{CNT} = \frac{2\gamma^\mathrm{eq}}{g_n^\mathrm{CNT}}
\end{equation*}
and
\begin{equation*}
    \Delta\varOmega_\mathrm{c}^\mathrm{CNT} = \frac{16\mathrm{\pi}{(\gamma^\mathrm{eq})}^3}{3(g_n^\mathrm{CNT})^2}.
\end{equation*}
As supersaturation increases, the size of the critical cluster and the work of formation initially decrease, then asymptotically approach non-zero values. 

At the binodal limit, our two-variable and three-variable models, as well as the full SGA model, recover the traditional predictions of CNT: the critical density equals the equilibrium liquid density, while both the critical radius and work of formation diverge. At low supersaturation ($\rho_0$ slightly above $\rho_1^\mathrm{eq}$), all four models predict very similar critical radii and works of formation, both of which decrease monotonically with increasing supersaturation. In contrast, their predictions for the critical density differ. CNT assumes a constant critical density equal to the equilibrium liquid density. Our two-variable model predicts a monotonic decrease of the critical density with supersaturation, whereas the three-variable model and the full SGA model both predict critical densities that first increase slightly then decrease as supersaturation increases. At intermediate supersaturation, our two-variable and three-variable models, as well as the full SGA model, predict substantially lower critical densities and nucleation barriers than CNT. Although the thermodynamic driving force for nucleating a dilute cluster is smaller than that for a cluster with the equilibrium liquid density, this reduction is more than compensated by a significantly lower surface energy cost, leading to a reduced work of formation. Upon approaching the spinodal limit, both of our models recover behavior consistent with the full SGA model: the critical cluster size diverges, its density approaches that of the initial metastable phase, and the work of formation vanishes. This behavior provides a smooth transition from nucleation to spinodal decomposition, where the energy barrier is zero, as established in diffuse-interface models \cite{Cahn-Hilliard-1959} and recovered in previous nucleation models with two parameters \cite{Schmelzer_2000, Schmelzer_2006, Schmelzer_2003, Schmelzer_2011, Baidakov_2000, Philippe_2011_Phil_mag, Philippe_2024, Bonvalet_Phil_mag_2014, Ghosh-2011, Reiss_1976, Schmelzer_2007}. CNT fails to capture this behavior, highlighting the improved physical accuracy of our two-variable and three-variable models. 

Our three-variable model yields critical density profiles in very close agreement with those obtained from the full SGA model, confirming the quality of the three-variable density profile ansatz. As a result, the predicted critical cluster density, radius ($\bar{R}$ but not $R$), and work of formation are nearly identical in the two approaches. Additionally, both models predict a progressive increase in interface diffuseness with supersaturation, which becomes particularly rapid at high supersaturation. 

Our two-variable model yet gives quite similar predictions compared to the full SGA model. Aside from the critical density, noticeable differences can also be seen in the critical work of formation. This difference is due to both the different descriptions of the density profile and the different ways of calculating the work of formation. 

We further compare the predictions of the various models with microscopic MD simulations. In Ref.~\onlinecite{Wu2026PRL}, only a limited number of critical configurations ($\sim 20$ or less) at a given gas density were obtained from rare-event sampling simulations. Because the radial density profiles exhibit large statistical fluctuations, the cluster properties were extracted by fitting the integral of the modified sigmoid function, Eq.~(\ref{eq:ModifiedSigmoid}), to the integrated MD density profiles. The latter were computed using a bin size of $1.0 \,\sigma$. Although this procedure revealed an overall decrease in the critical cluster density with increasing supersaturation, it did not reconcile the diffuse-interface prediction for the core density. Two factors may account for this discrepancy. First, the relatively large bin size may smear the diffuse interface into the cluster core, leading to an underestimate of the core density. Second, fitting the integrated density profile may not be sufficiently sensitive to resolve the behavior predicted by the full SGA model.

To examine the structure of critical clusters in greater detail over the supersaturation range of interest, we perform new seeded MD simulations in the \textit{NVT} ensemble~\cite{Vega-Seeding-NVT,PHILIPPE2026108254}. A large collection of critical clusters is generated, allowing the radial density profiles to be averaged over thousands of critical configurations with the same gas density $\rho_0$. The modified sigmoid function, Eq.~(\ref{eq:ModifiedSigmoid}), is then fitted directly to the averaged radial density profile to determine the critical radius $R_\mathrm{c}$, the core density $\rho_\mathrm{c}$, the interfacial width $w_\mathrm{c}$, and the surrounding gas density $\rho_0$. To assess the robustness of the fitting procedure, the analysis is repeated using bin sizes between $0.5$ and $1.0 \,\sigma$, yielding nearly identical results. In particular, the non-monotonic dependence of $\rho_\mathrm{c}$ on supersaturation is consistently recovered for all bin sizes. 

\begin{figure}[th]
\begin{tikzpicture}
\node (fig1) at (0,0) {\includegraphics[trim={0.3cm 0cm 0.3cm 0cm},clip,width=0.45\textwidth]{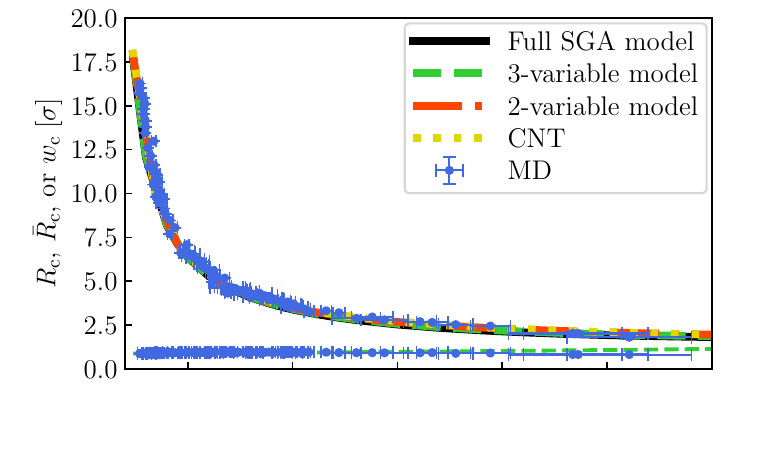}};
\node[] at (-4,2.5) {(a)};
\node (fig2) at (0,-4.13) {\includegraphics[trim={0.3cm 0cm 0.3cm 0cm},clip,width=0.45\textwidth]{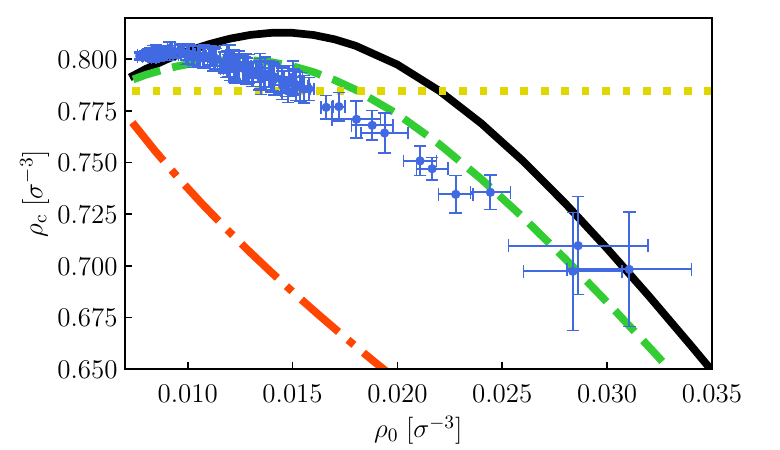}};
\node[] at (-4,-1.8) {(b)};
\end{tikzpicture}
\caption{Comparison of the (a) radius (as well as interfacial width) and (b) density of the critical cluster predicted by various models with our MD simulation results, at $T=0.8\epsilon/k_\mathrm{B}$. Our MD simulation results are represented by blue circles with error bars indicating standard deviations. Predictions from our two-variable model, three-variable model, CNT, and the full SGA model are represented by red dash-dotted lines, green dashed lines, yellow dotted lines, and black solid lines, respectively. The thinner green dashed line in panel (a) represents the interfacial width of the critical cluster predicted by the three-variable model. }
\label{fig:R_c and rho_c with MD}
\end{figure}

The radius, interfacial width, and core density of the critical clusters extracted from our seeded simulations, together with predictions of different models, are shown in Fig.~\ref{fig:R_c and rho_c with MD}. Over the range of supersaturation investigated ($\rho_0<0.035 \,\sigma^{-3}$), all models predict nearly identical critical radii, in excellent agreement with the simulation results. The critical interfacial width remains nearly constant at $\sim 1.0\,\sigma$ and is precisely reproduced by the three-variable model. The situation is markedly different for the critical cluster density. Our MD simulations indicate that the critical core density exhibits a weak increase at very low supersaturation, reflecting the Gibbs--Thomson effect, followed by a decrease as the supersaturation increases. This non-monotonic behavior is consistent with earlier predictions from other multidimensional nucleation theories \cite{Reiss_1976,Schmelzer_2007,Philippe_2011_Phil_mag} and diffuse-interface approaches \cite{Cahn-Hilliard-1959,Philippe_JCP_2011,Lutsko-JCP2011,Simeone_PRL2023}. Our three-variable model and the full SGA model successfully reproduce this trend. In contrast, CNT fails to capture this behavior. Our two-variable model reproduces only the decreasing trend and systematically underestimates the critical density, reflecting the limitations imposed by its simplified sharp-interface description of the cluster density profile.

\subsection{Nucleation rate and its pre-factors}

\begin{figure}[tbh]
\begin{tikzpicture}
\node (fig1) at (0,0) {\includegraphics[trim={0.3cm 0cm 0.3cm 0cm},clip,width=0.45\textwidth]{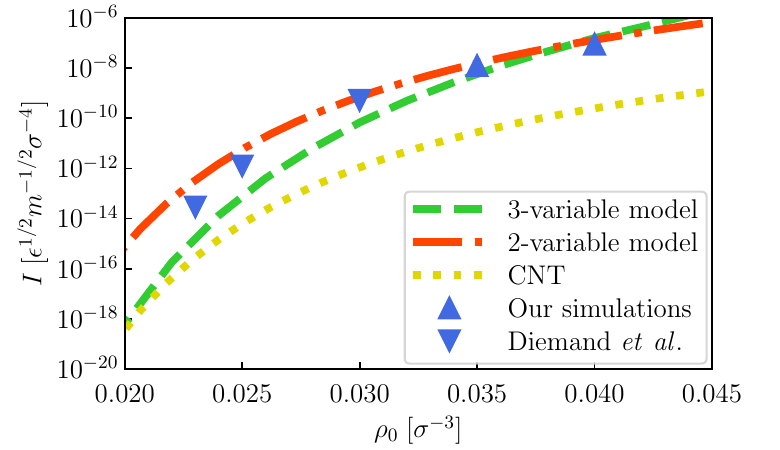}};
\end{tikzpicture}
\caption{Evolutions of the steady-state nucleation rate as functions of the initial density at $T=0.8\epsilon/k_\mathrm{B}$. Our brute-force simulation results~\cite{Wu2026PRL} are represented by blue up-pointing triangles, and results from Diemand \textit{et al.}~\cite{Diemand_JCP2013} are represented by blue down-pointing triangles. Predictions from our two-variable model, three-variable model, and CNT are represented by red dash-dotted lines, green dashed lines, and yellow dotted lines, respectively. }
\label{fig:I}
\end{figure}

Next, we use our two-variable and three-variable models to predict the steady-state nucleation rate. We emphasize that these models contain no adjustable parameters. The predicted nucleation rates, based on the DFT database, are reported in Fig.~\ref{fig:I}. For comparison, we also show the CNT prediction, 
\begin{equation*}
    I^\mathrm{CNT} =  f_0^\mathrm{CNT} \frac{|{\lambda_\mathbf{Z}}_1^\mathrm{CNT}|}{4\mathrm{\pi}\sqrt{k_\mathrm{B}T\gamma^\mathrm{eq}}} \exp\big(- \frac{\Delta\varOmega_\mathrm{c}^\mathrm{CNT}}{k_\mathrm{B}T}\big) ,
\end{equation*}
where the negative eigenvalue of the matrix $\mathbf{Z}$ in CNT reads 
\begin{equation}
\label{eq:lambda_CNT}
    {\lambda_\mathbf{Z}}_1^\mathrm{CNT} = -\frac{D_\mathrm{diff}\, \rho_0 {(g_n^\mathrm{CNT})}^3}{4(\rho_2^\mathrm{eq}- \rho_0)^2 {\gamma^\mathrm{eq}}^2} ,
\end{equation}
and the distribution constant, computed from mass conservation, is 
\begin{equation*}
    f_0^\mathrm{CNT} = \dfrac{\rho_0} {\frac{4}{3}\mathrm{\pi} \rho_\mathrm{c}^\mathrm{CNT} \displaystyle\int_0^{R_\mathrm{c}^\mathrm{CNT}} R^3 \exp{\big(-\frac{\Delta\varOmega^\mathrm{CNT}(R)}{k_\mathrm{B}T}\big) \,\mathrm{d}R}} .
\end{equation*}
Here, the CNT work of formation is $\Delta\varOmega^\mathrm{CNT}(R)=-4\mathrm{\pi}{R}^3 g_n^\mathrm{CNT}/3 + 4\mathrm{\pi}{R}^2\gamma^\mathrm{eq}$. 

Our predicted nucleation rates are compared with brute-force MD simulations, either performed by ourselves \cite{Wu2026PRL} or reported in Ref.~\onlinecite{Diemand_JCP2013}. Remarkably, our two-variable model performs very well over the supersaturation range accessible to the MD simulations. Our three-variable model also shows good overall agreement with the simulation results, whereas CNT underestimates the rate by several orders of magnitude. The differences among the predictions of the various models primarily originate from the critical work of formation, which enters the nucleation rate exponentially. Over the range investigated~($0.02 \,\sigma^{-3} \leq \rho_0 \leq 0.04 \,\sigma^{-3}$), compared to both the three-variable and the full SGA models, the two-variable model predicts a slightly smaller critical work of formation, whereas CNT predicts a larger one, as shown in Fig.~\ref{fig:R_c,rho_c,DeltaOmega_c}(c). 

\begin{figure}[htb]
\begin{tikzpicture}
\node (fig1) at (0,0) {\includegraphics[trim={0.3cm 0cm 0.3cm 0cm},clip,width=0.45\textwidth]{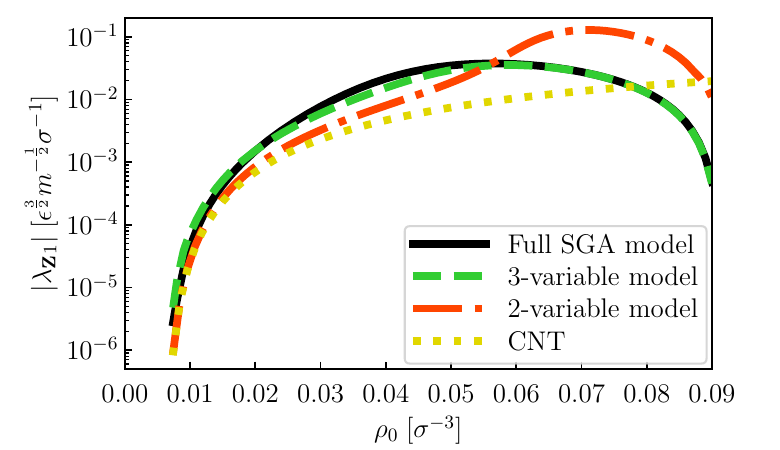}};
\end{tikzpicture}
\caption{Evolutions of the absolute value of the negative eigenvalue of $\mathbf{Z}$ as functions of the initial density at $T=0.8\epsilon/k_\mathrm{B}$. Predictions from our two-variable model, three-variable model, CNT, and the full SGA model are represented by red dash-dotted lines, green dashed lines, yellow dotted lines, and black solid lines, respectively.}
\label{fig:lambdaZ1}
\end{figure}

Computing the nucleation rate, or even its prefactor, within the full SGA model is notoriously challenging because of the complexity on the determination of $f_0$. For this reason, such a calculation is not pursued here. Nevertheless, the negative eigenvalue ${\lambda_\mathbf{Z}}_1$ can still be evaluated and meaningfully compared with those obtained from other models. Predictions of the absolute value of ${\lambda_\mathbf{Z}}_1$ from various models are shown in Fig.~\ref{fig:lambdaZ1}. 

Since the growth of the system at the very early stage just after nucleation proceeds along the unstable direction of the system, i.e., the nucleation flux direction, the eigenvalue ${\lambda_\mathbf{Z}}_1$ is invariant under any reparameterization that preserves the same unstable direction. A detailed explanation of this invariance is provided in Appendix~\ref{Appendix:lambdaZ}. At low supersaturation, the values of ${\lambda_\mathbf{Z}}_1$ predicted by all models are nearly identical. This indicates that, in this regime, cluster size is the very dominant unstable variable in multivariable models, in full consistency with the nucleation pathway analyses presented later. 
Clear discrepancies emerge at intermediate supersaturation. All models except CNT exhibit a nonmonotonic dependence of ${\lambda_\mathbf{Z}}_1$ on supersaturation. As soon as ${\lambda_\mathbf{Z}}_1$ deviates from the CNT prediction, the nucleation process can no longer be adequately described using cluster size as the sole reaction coordinate, signaling the breakdown of the classical picture. More generally, the predictions of our three-variable model closely track those of the full SGA model. This agreement further validates the three-variable ansatz for the density profile. 
\revision{The three-variable model therefore provides a practical alternative to the full SGA model because it enables calculation of the nucleation rate. Moreover, reproducing the full SGA predictions with only three collective variables provides additional physical insight: cluster size, density, and interfacial width capture the dominant features of the condensation process.}

\subsection{Nucleation pathway}
Finally, we analyze the nucleation flux direction. Within CNT, both the nucleation flux and energy descent directions coincide with the direction of increasing cluster size. In contrast, our two- and three-variable models reveal more intricate nucleation pathways. 

On the one hand, we compare the nucleation flux direction with the energy steepest descent direction. As discussed earlier, the nucleation flux direction is given by the eigenvector of the matrix $\mathbf{Z}$ that corresponds to its unique negative eigenvalue ${\lambda_\mathbf{Z}}_1$, denoted by ${\boldsymbol{v}_\mathbf{Z}}_1$. The energy steepest descent direction corresponds to the sole descending principal axis of the work-of-formation surface toward the stable region. For clarity, we present results from the two-variable model only. 

\begin{figure}[tbh]
\centering
\begin{tikzpicture}
\node (fig1) at (0,0) {\includegraphics[trim={0.3cm 0cm 0.3cm 1.9cm},clip,width=0.45\textwidth]{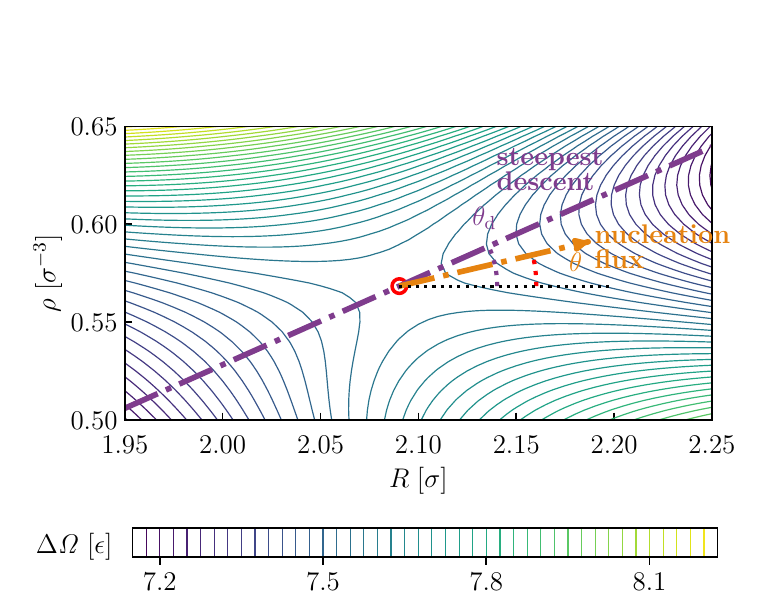}};
\node[] at (-4,2.5) {(a)};
\node (fig2) at (0,-5.7) {\includegraphics[trim={0.3cm 0cm 0.3cm 0cmcm},clip,width=0.45\textwidth]{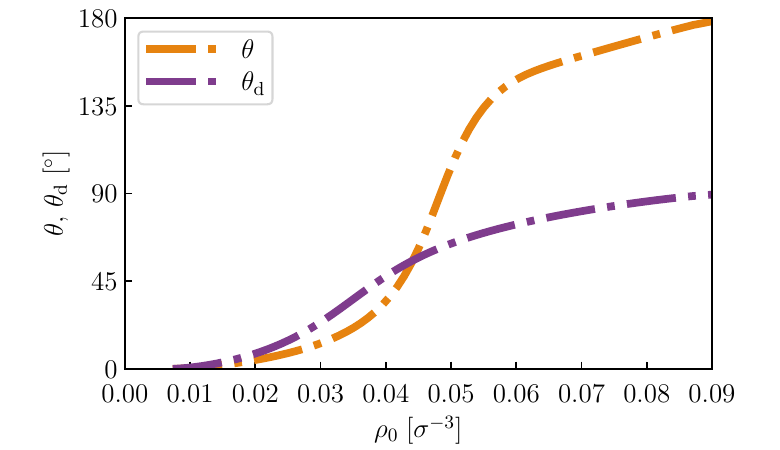}};
\node[] at (-4,-3.4) {(b)};
\end{tikzpicture}
\caption{(a) Contour plot of the work of formation with nucleation flux and energy steepest descent directions at $T=0.8\epsilon/k_\mathrm{B}$ and $\rho_0=0.03\sigma^{-3}$. The saddle point is marked by a red circle. The orange and violet dash-dotted lines indicate the predicted nucleation flux and steepest descent directions at the saddle point, respectively. (b) Evolution of nucleation flux and steepest descent directions against the initial density at $T=0.8\epsilon/k_\mathrm{B}$. Predictions are from our two-variable model.}
\label{fig:theta}
\end{figure}

Fig.~\ref{fig:theta}(a) shows contour lines of work of formation in the $R$-$\rho$ phase space for $\rho_0 = 0.03 \,\sigma^{-3}$. The saddle point is indicated by the red circle. The energy steepest descent direction is represented by the violet dash-dotted line, forming an angle $\theta_\mathrm{d}$ with the $+R$ direction. The nucleation flux direction is shown by the orange dash-dotted arrow, with its angle relative to the $+R$ direction denoted by $\theta$. In the very early stage of growth, trajectories extracted from our MD simulations exhibit concomitant cluster growth and densification, and align more closely with the nucleation flux direction predicted by our two-variable model than with its corresponding energy steepest descent direction, as shown in our previous publication \cite{Wu2026PRL}. 
Fig.~\ref{fig:theta}(b) displays the evolution of steepest descent and nucleation flux directions against the initial density. At low supersaturation, both directions coincide with the $+R$ direction. As supersaturation increases, they rotate counter-clockwise. Near the spinodal limit, the steepest descent direction aligns with the $+\rho$ direction, whereas the nucleation flux direction points toward the $-R$ direction. These results indicate nucleation pathways that differ qualitatively from those predicted by CNT. 

\begin{figure}[th]
\begin{tikzpicture}
\node (fig1) at (0,0) {\includegraphics[trim={0.3cm 0cm 0.3cm 0cm},clip,width=0.45\textwidth]{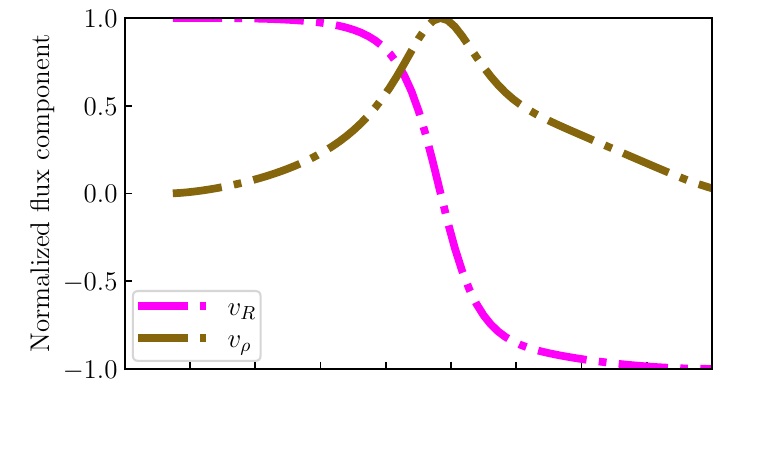}};
\node[] at (-4,2.5) {(a)};
\node (fig2) at (0,-4.3) {\includegraphics[trim={0.3cm 0cm 0.3cm 0cm},clip,width=0.45\textwidth]{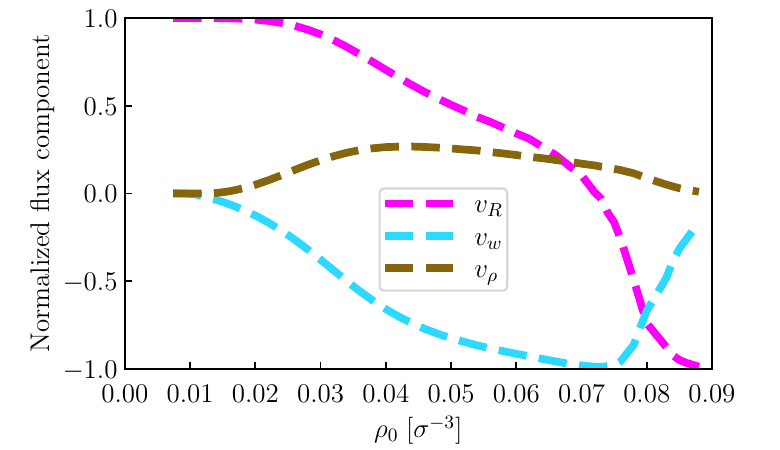}};
\node[] at (-4,-2) {(b)};
\end{tikzpicture}
\caption{Evolutions of the normalized nucleation flux components against the initial density at $T=0.8\epsilon/k_\mathrm{B}$, predicted by our (a) two-variable nucleation model and (b) three-variable nucleation model. The $R$, $\rho$, and $w$ components, which are nondimensionalized by Lennard-Jones reduced units, are shown in magenta, brown, and cyan, respectively; the $w$ component appears only in the three-variable model. }
\label{fig:NucleationFluxElements}
\end{figure}

On the other hand, we compare the nucleation flux directions predicted by our two- and three-variable nucleation models. The flux direction is represented by a dimensionless unit vector in the corresponding order-parameter space, namely ${\boldsymbol{v}_\mathbf{Z}}_1 = (v_R,v_\rho)^\mathrm{T}$ in the two-variable $(R,\rho)$ space and ${\boldsymbol{v}_\mathbf{Z}}_1 = (v_R,v_w,v_\rho)^\mathrm{T}$ in the three-variable $(R,w,\rho)$ space. The results are shown in Fig.~\ref{fig:NucleationFluxElements}. 

Regarding the $R$ component, both models predict a similar trend: $v_R$ evolves continuously from $+1$ to $-1$. Near the binodal limit, the nucleation pathway follows the classical picture, with the flux directed toward increasing cluster size. In contrast, close to the spinodal limit, the nucleation flux points toward decreasing cluster size, indicating that the critical cluster initially shrinks with only tiny changes in the other variable(s). Although this behavior may appear counterintuitive, we emphasize that the nucleation flux direction characterizes only the very first stage of deterministic growth and should not be interpreted as the full postcritical growth trajectory. Moreover, the work of formation becomes very small when approaching the spinodal region, and saddle-point nucleation becomes questionable since ridge-crossing, provoked by thermal fluctuations, may intervene.

The remaining component(s) of the flux vector must evolve concurrently to compensate for the change in $v_R$. In our two-variable model, this role is played solely by $v_\rho$. In our three-variable model, however, both $v_w$ and $v_\rho$ contribute, with $v_w$ being the dominant component. The signs $v_\rho > 0$ and $v_w < 0$ further imply that, at the very early stage of postcritical growth, the cluster density increases while the interface becomes sharper.

\section{Conclusion}
\label{sec:conclusion}
    By extending CNT to include both cluster size and density as independent variables within the capillary approximation, and by further incorporating interface diffuseness through a three-variable diffuse-interface model, we have revealed the complex interplay between kinetics and thermodynamics that gives rise to nonclassical nucleation mechanisms. While the classical picture is recovered near the binodal limit, it progressively breaks down at intermediate and high supersaturation, where cluster size, density, and, within the diffuse-interface description, interfacial width evolve concomitantly. 

Our two-variable and three-variable nucleation models are fully self-consistent, with all parameters determined from physical considerations. The models were systematically validated against MD simulations of Lennard-Jones condensation. In the low and intermediate supersaturation regimes, both models accurately predict the critical cluster size. For the critical cluster density, however, only the three-variable diffuse-interface model yields accurate predictions and correctly captures the weak non-monotonic dependence of the critical cluster density at very low supersaturation. In contrast, the two-variable model reproduces only the overall decreasing trend of the critical density. In the regime of supersaturation accessible with brute-force simulations, both models were further validated through their quantitative predictions of the nucleation rate. Moreover, compared with other theoretical approaches, both models substantially outperform CNT, while the three-variable model closely reproduces the critical cluster density profile, critical work of formation, and system growth rate obtained from the Euler--Lagrange solution of the full SGA diffuse-interface model, establishing the three-variable model as a compelling and efficient alternative to the full SGA model. 
Regarding the nucleation pathway, both our models successfully capture the concomitant growth and densification of postcritical clusters. The three-variable model further provides a more physically consistent description of the nucleation pathway by allowing the interface width to evolve dynamically. 

More broadly, we anticipate that multivariable nucleation models coupled with dynamical density functional theory will provide a powerful framework for investigating complex nucleation scenarios, such as multicomponent or two-step nucleation processes in dilute systems that may involve multiple reaction coordinates \cite{2S-binary-LJ2023,Iwamatsu2011}. Identifying the appropriate reaction coordinates that govern nucleation pathways remains a central challenge and an open question in the study of first-order phase transitions.

\begin{acknowledgments}
This work was supported by the ANR TITANS project: ANR-21-CE06-0032. The authors thank Julien Lam (UMET, Lille) and Aymane Graini (LMS, Ecole Polytechnique) for helpful discussions.
\end{acknowledgments}

\section*{Data Availability Statement}
The data that support the findings of this study are available from the corresponding author upon reasonable request.

\appendix
\section*{Appendices}
\numberwithin{equation}{section}
\section{Consistency between Langer's and Alekseechkin's methods}
\label{Appendix:Langer&Alekseechkin}
\normalsize Langer's derivation \cite{langer_statistical_1969} starts from the following Fokker--Planck equation
\begin{equation}
    \pypx{f}{t} = -\boldsymbol{\nabla\cdot}\boldsymbol{J} , \quad 
    \boldsymbol{J} = -k_\mathrm{B}T \,\mathbf{M}\boldsymbol{\nabla}f + \pypx{\boldsymbol{x}}{t}f ,
\label{eq:FP_Langer}
\end{equation}
\revision{where $f=f(\boldsymbol{x},t)$ is the time-dependent distribution function of clusters in the $n$-dimensional phase space $\{\boldsymbol{x}\}$ (collective-variable space).}
$\mathbf{M}$ is the mobility matrix. Comparing this to Eq. (\ref{eq:FP_Alekseechkin}), we see that 
\begin{equation}
    \mathbf{D} = k_\mathrm{B}T\,\mathbf{M} .
\label{eq:DM_relation}
\end{equation}

A new frame of coordinates, $\{\boldsymbol{\xi}\}$, is established in the phase space via the orthogonal transformation (after nondimensionalization). It is the principal-axis coordinate of the work of formation with its origin at the saddle point, $\boldsymbol{x}_\mathrm{c}$. The transformation between the two coordinate systems is given by
\begin{equation}
    \boldsymbol{\xi} = \mathbf{C}(\boldsymbol{x} - \boldsymbol{x}_\mathrm{c}) .
\end{equation}
In the $\{\boldsymbol{\xi}\}$ system, matrices are denoted with $\widehat{\;}\;$ and transformed as follows: the Hessian matrix becomes diagonal, i.e., $\mathbf{\widehat{H}}=\mathbf{({C^\mathrm{-1}})^\mathrm{T} H C^\mathrm{-1}}$, and the mobility matrix becomes $\mathbf{\widehat{M}}=\mathbf{C M C^\mathrm{T}}$. 

Langer's general expression for the steady-state nucleation theory is 
\begin{equation}
    I=f_0\cdot \exp{\big(-\dfrac{\Delta\varOmega_\mathrm{c}}{k_\mathrm{B}T}\big)} \cdot |K| \cdot \sqrt{\frac{ k_\mathrm{B}T}{2\picst |\lambda_1|}} \cdot {\prod \limits_n}^{''}\sqrt{\frac{2\picst k_\mathrm{B}T}{\lambda_i}} .
\label{eq:I_Langer}
\end{equation}
Here, $K$ is the only negative eigenvalue of the matrix $\mathbf{\widehat{L}}=\mathbf{\widehat{H}\widehat{M}}$, and $\lambda_i$'s are the eigenvalues of $\mathbf{H}$ (or equivalently $\mathbf{\widehat{H}}$ in the principal-axis coordinate system), with $i=1$ corresponding to the only negative one. The notation ${\prod \limits_n}^{''}$ excludes terms where $\lambda_i \leq 0$. Assuming only non-zero eigenvalues of $\mathbf{H}$, as done in Alekseechkin's work \cite{alekseechkin_multivariable_2006}, Eq. (\ref{eq:I_Langer}) simplifies to
\begin{equation}
    I=f_0\cdot \exp{\big(-\dfrac{\Delta\varOmega_\mathrm{c}}{k_\mathrm{B}T}\big)} \cdot \big(2\picst k_\mathrm{B}T\big)^{\frac{n-2}{2}} \cdot \frac{k_\mathrm{B}T|K|}{\sqrt{|\det \mathbf{H}|}} .
\label{eq:I_Langer_arranged}
\end{equation}

The nucleation flux direction in the $\{\boldsymbol{x}\}$ coordinate is $\mathbf{C}^\mathrm{-1}(U_1/\lambda_1, U_2/\lambda_2, \dots, U_n/\lambda_n)^\mathrm{T}$, where $(U_1, U_2, \dots, U_n)^\mathrm{T}$ is the eigenvector of $\mathbf{\widehat{L}}$ corresponding to $K$ in the $\{\boldsymbol{\xi}\}$ coordinate. 

On the other hand, Alekseechkin's expression for the steady-state nucleation rate, given by Eq. (\ref{eq:I}), corresponds to the Fokker--Planck equation in Eq. (\ref{eq:FP_Alekseechkin}). In his framework, the nucleation flux direction is along the eigenvector of the matrix $\mathbf{Z}$ corresponding to its only negative eigenvalue, ${\lambda_\mathbf{Z}}_1$, denoted as ${\boldsymbol{v}_\mathbf{Z}}_1$, in the $\{\boldsymbol{x}\}$ coordinate system.

The two methods, Langer's and Alekseechkin's, are equivalent (except for the equilibrium distribution constant, $f_0$) if the following conditions hold: 
\begin{equation*}
    {\lambda_\mathbf{Z}}_1=k_\mathrm{B}TK
\end{equation*}
and 
\begin{equation*}
    \mathbf{C}^\mathrm{-1}(U_1/\lambda_1, U_2/\lambda_2, \dots, U_n/\lambda_n)^\mathrm{T} = c {\boldsymbol{v}_\mathbf{Z}}_1,\,c\neq0 .
\end{equation*}
This equivalence can be demonstrated as follows.

From the transformations mentioned above, we can write 
\begin{equation*}
    \mathbf{\widehat{L}} = \mathbf{\widehat{H}\widehat{M}} = \mathbf{({C^\mathrm{-1}})^\mathrm{T} H C^\mathrm{-1}} \mathbf{C M C^\mathrm{T}} = \mathbf{({C^\mathrm{-1}})^\mathrm{T} H M C^\mathrm{T}} .
\end{equation*}
Thus, the mobility matrix can be expressed as 
$\mathbf{M = H^\mathrm{-1} C^\mathrm{T} \widehat{L} ({C^\mathrm{-1}})^\mathrm{T}}$. Substituting this into Eq. (\ref{eq:DM_relation}), the matrix $\mathbf{Z}$ becomes
\begin{equation}
    \mathbf{Z} \equiv \mathbf{DH} = k_\mathrm{B}T\,\mathbf{MH} = k_\mathrm{B}T\, \mathbf{H^\mathrm{-1} C^\mathrm{T} \widehat{L} ({C^\mathrm{-1}})^\mathrm{T} H} .
\end{equation}
Thus, $\mathbf{Z}$ is similar to $k_\mathrm{B}T\,\mathbf{\widehat{L}}$. Since similar matrices share the same eigenvalues and their eigenvectors are relevant, it follows that  
\begin{equation}
    {\lambda_\mathbf{Z}}_1=k_\mathrm{B}TK
\end{equation}
and 
\begin{equation}
    c{\boldsymbol{v}_\mathbf{Z}}_1 = \mathbf{{H^\mathrm{-1}} C^\mathrm{T}}  (U_1, U_2, \dots, U_n)^\mathrm{T} ,\quad c\neq 0 .
\label{eq:eigenvectorZL}
\end{equation}
In the principal-axis coordinate system, the Hessian matrix $\mathbf{\widehat{H}}$ is diagonal,
\begin{equation*}
    \mathbf{\widehat{H}} = 
    \begin{pmatrix}
        \lambda_1 &  &   \\
         & \ddots &    \\
          & & \lambda_n
    \end{pmatrix}
    ,
\end{equation*}
so its inverse, $\mathbf{{\widehat{H}}^\mathrm{-1}}$, is also diagonal,
\begin{equation*}
    \mathbf{{\widehat{H}}^\mathrm{-1}} = 
    \begin{pmatrix}
        \dfrac{1}{\lambda_1} &  &   \\
         & \ddots &    \\
          & & \dfrac{1}{\lambda_n}
    \end{pmatrix}
    = \mathbf{C H^\mathrm{-1} C^T} .
\end{equation*}
Finally, Eq. (\ref{eq:eigenvectorZL}) becomes
\begin{eqnarray}
    c{\boldsymbol{v}_\mathbf{Z}}_1 &&= \mathbf{C^\mathrm{-1} {\widehat{H}}^\mathrm{-1}} (U_1, U_2, \dots, U_n)^\mathrm{T} \nonumber  \\
    &&= \mathbf{C}^\mathrm{-1}(U_1/\lambda_1, U_2/\lambda_2, \dots, U_n/\lambda_n)^\mathrm{T} .
\end{eqnarray}
This confirms the expected relationship between Langer's and Alekseechkin's methods.

\section{Normalization constant}
\label{Appendix:f0}
The constant $f_0$ in the equilibrium density function of nuclei, 
\begin{equation*}
    f_\mathrm{eq} = f_0\,\exp{\big(-\frac{\Delta\varOmega(\boldsymbol{x})}{\kBT}\big)}
\end{equation*}
is a key factor in determining the steady-state nucleation rate, as given in Eq. (\ref{eq:I}). Here, we discuss its determination proposed by Alekseechkin and Langer. 

Alekseechkin \cite{alekseechkin_multivariable_2006} calculated $f_0$ by expanding the work of formation at the saddle point and assuming that $x_1$ represents the number of monomers in the cluster, $H_{11}<0$ and $H_{ii}>0$ for any $i>1$. Then $f_0 = \rho_0\sqrt{\mathrm{H_{11}^{-1}}\cdot \det \mathbf{H}}\cdot(2\picst\kBT)^{-\frac{n-1}{2}}$, with $\mathrm{H_{11}^{-1}}$ the $(1,1)$ element of the inverse matrix of $\mathbf{H}$. It should be noted that in his paper, $\mathrm{H_{11}}$ has an energy unit. In a more general case, $x_1$ can be any variable representing the size of the cluster, then 
\begin{equation}
    f_0 = f^{x_1}_0\sqrt{\mathrm{H_{11}^{-1}}\cdot \det \mathbf{H}}\cdot(2\picst\kBT)^{-\frac{n-1}{2}} ,
\label{eq:f0_Alekseechkin}
\end{equation}
where $\displaystyle\int_0^{{x_1}_\mathrm{c}}f^{x_1}_0\exp{\big(-\frac{\Delta\varOmega}{k_\mathrm{B}T}\big)}\mathrm{d}x_1 = \rho_0$ and ${x_1}_\mathrm{c}$ is the corresponding value at critical size. However, this expansion at the saddle point fails to capture the behavior of the work of formation for clusters near the initial state, which is critical for determining the Boltzmann distribution. Additionally, our studies show that the assumption on the diagonal elements of $\mathbf{H}$ is not universally valid. 

Langer \cite{langer_statistical_1969} handled the problem by integrating the distribution of clusters in the subcritical side and normalizing the function $f$ as a probability density function, 
\begin{equation}
    \int_\mathrm{subC} f_\mathrm{L}(\boldsymbol{x}) \mathrm{d}\boldsymbol{x} = 1 ,
\end{equation}
which gave his steady-state nucleation rate the interpretation of probability per unit time. Then, he expanded the work of formation at the metastable minimum, i.e., the initial state, assuming that the metastable minimum is sharp and well isolated, which means the eigenvalues of Hessian matrix of work of formation at this metastable minimum, $\mathbf{H}_0$, are all non-zero. This assumption, however, does not hold in our studies, either. In our two-variable model, both eigenvalues of $\mathbf{H}_0$ are zero (in fact, $\mathbf{H}_0$ is just a zero matrix). In our three-variable model, the partial derivatives of work of formation with respect to 
$R$ and $w$ at $\rho = \rho_0$ are always zero, so $\mathbf{H}_0$ definitively has a zero eigenvalue. As a result, Langer's formulation for $f_0$ is inapplicable to our studies.

If we adhere to Langer's integral approach and adjust it by incorporating the density of clusters to ensure that the nucleation rate is expressed in units of per volume per time, we arrive at Eq. (\ref{eq:f0_normalization}). However, the number density of clusters is unknown.

\section{Recipe for calculating Helmholtz free energy via perturbation theory}
\label{Appendix:Recipe F}
Thermodynamic perturbation theory is used to calculate the Helmholtz free energy, $F$, of a system of $N$ particles. This is achieved by treating the system as a simple system with an added perturbation. Various approaches exist for constructing the perturbation, with the best known being the Weeks--Chandler--Andersen (WCA) theory \cite{chandler_equilibrium_1970,weeks_role_1971,andersen_relationship_1971}. We follow the procedure in the paper of Lutsko and Nicolis \cite{lutsko_effect_2005} for computing the bulk Helmholtz free energy of a system. 

The interaction potential of the system, $v(r)$, is first decomposed into a repulsive short-ranged part, $v_0(r)$, and a (mostly) attractive longer-ranged tail, $w(r)$, as per the modified WCA theory \cite{ree_equilibrium_1976,kang_perturbation_1985},
\begin{equation*}
\begin{aligned}
     &v_0(r) = \left\{
    \begin{aligned}
        &v(r)-v(r_0)-v'(r_0)(r-r_0), \quad r<r_0    \\[7pt]
        &0, \quad r\geq r_0
    \end{aligned}
    \right. \\[7pt]
    &w(r) = \left\{
    \begin{aligned}
        &v(r_0)+v'(r_0)(r-r_0), \quad r<r_0    \\[7pt]
        &v(r), \quad r\geq r_0
    \end{aligned}
    \right.
\end{aligned}
.
\end{equation*}
The separation point $r_0$ is calculated based on Ref.~\onlinecite{kang_perturbation_1986},

\begin{equation}
    r_0 = r_\mathrm{min}+S(r_\mathrm{nn}-r_\mathrm{min}),
\end{equation}
with $r_\mathrm{min}$ the minimum of the potential, defined by $v'(r_\mathrm{min})=0$, and $r_\mathrm{nn}(\rho)=2^{1/6}\rho^{-1/3}$ the nearest-neighbor distance in the fcc lattice. $S$ is a function of the density $\rho$, defined piecewise as
\begin{widetext}
\begin{equation*}
    S = \left\{
    \begin{aligned}
        &0, \quad \rho\leq\rho_1    \\[7pt]
        &\frac{(\rho-\rho_1)^3(6\rho^2-3(5\rho_2-\rho_1)\rho+10\rho_2^2-5\rho_1\rho_2+\rho_1^2)}{(\rho_2-\rho_1)^5}, \quad \rho_1<\rho<\rho_2   \\[7pt]
        &1, \quad \rho\geq \rho_2
    \end{aligned}
    \right. 
    ,
\end{equation*}
\end{widetext}
where $\rho_1=0.97\rho_\mathrm{s}$ and $\rho_2=1.01\rho_\mathrm{s}$, with $\rho_\mathrm{s}$ being the density at which $r_\mathrm{nn}(\rho_\mathrm{s})=r_\mathrm{min}$.

Given this decomposition, the Helmholtz free energy of the system can be expanded in the high temperature series \cite{hansen_theory_2013}, 
\begin{equation*}
    \frac{1}{N}\beta F = \frac{1}{N}\beta F_0 + \frac{1}{N}\beta \langle W\rangle_0 - \frac{1}{2N}\beta^2 \Big(\langle W^2\rangle_0 - \langle W\rangle_0^2\Big) + \cdots .
\end{equation*}
Here, $F_0$ is the free energy of the reference system subject to the short-ranged potential $v_0(r)$ at inverse temperature $\beta=1/(\kBT)$, and other terms on the right-hand side (RHS) with $W$ are components of the free energy of the perturbation. $W$ represents the total energy subject to the longer-ranged potential $w(r)$, and the angle brackets with subscript, $\langle \, \rangle_0$, indicate an equilibrium average over a system interacting with the potential $v_0$. The first-order term of the perturbation can be calculated from the pair distribution function of the reference system, $g_0(r)$, as
\begin{equation}
    \frac{1}{N}\beta \langle W\rangle_0 = \frac{\beta\rho}{2}\int g_0(r)w(r)\,\mathrm{d}\mathbf{r} .
\label{eq:1st PerturbationTerm}
\end{equation}
The $n^\mathrm{th}$-order term requires the distribution functions of all orders up to $2n$. Good approximations of these distribution functions are unavailable, while fortunately the higher-order terms become negligible compared with the first-order term if the decomposition of the potential is well-chosen \cite{hansen_theory_2013,lutsko_effect_2005}.  

The reference system, often referred to as the ``soft-core system", can be approximated by a hard-sphere system. For a harshly repulsive but continuous reference system \cite{hansen_theory_2013}, its free energy can be expressed as
\begin{equation}
    \frac{1}{N} \beta F_0 = 
    \frac{1}{N} \beta F_\mathrm{hs} + \frac{1}{N} \beta \displaystyle\int \left. \frac{\delta F_0}{\delta e_0 (\mathbf{r})} \right|_{e_0 = e_d} \Delta e(\mathbf{r}) \, \mathrm{d}\mathbf{r} + \cdots ,
\end{equation}
where $e_0 (r) = \exp(-\beta v_0 (r))$ is the Boltzmann factor of the reference system, and $e_{d} (r) = \exp(-\beta v_\mathrm{hd} (r; d))$ is the Boltzmann factor of the hard-sphere system with diameter $d$, and $\Delta e(r) = e_0(r) - e_{d}(r)$. Higher-order terms are negligible for steep potentials. 

The functional derivative can be calculated as
\begin{equation*}
    \left. \frac{\delta F_0}{\delta e_0 (\mathbf{r})} \right|_{e_0 = e_d} = -\frac{N}{2 \beta} \rho y_d (\mathbf{r}), 
\end{equation*}
with $y_d (r) = g_d (r) \exp(\beta v_\mathrm{hd} (r; d)) = g_d (r)/e_d (r)$ the hard-sphere cavity function and $g_d(r)$ the pair distribution function of the hard-sphere system. Then
\begin{equation}
    \frac{1}{N} \beta F_0 = \frac{1}{N} \beta F_\mathrm{hs} - \frac{1}{2} \rho \int y_d (\mathbf{r}) \Delta e(\mathbf{r}) \, \mathrm{d}\mathbf{r} + \cdots .
\end{equation}

The next goal is to find a proper effective hard-sphere diameter. We use the commonly used Barker and Henderson approximation \cite{hansen_theory_2013}, 
\begin{equation}
    d(T) = \int_0^\infty \big(1 - \exp(-\beta v_0 (r))\big) \, \mathrm{d} r .
\label{eq:d_hs}
\end{equation}
We have tested that this approximation yields very accurate results except at high density, compared to a more complicated but precise calculation of $d(T,\rho)$, which is to satisfy
\begin{equation*}
    \int y_d (\mathbf{r}) \Delta e(\mathbf{r}) \, \mathrm{d}\mathbf{r} = 0 .
\end{equation*}

The pair distribution function of the reference system, $g_0(r)$, can be approximated by $g_0(r)\approx g_d(r)$, so the final formula of the free energy of the original system becomes 
\begin{equation}
    \frac{1}{N}\beta F \approx \frac{1}{N} \beta F_\mathrm{hs} + 2\picst\beta\rho\int r^2g_d(r)w(r)\,\mathrm{d}r .
\label{eq:F}
\end{equation}

Its first term, the free energy of the hard-sphere system, can be computed via the Carnahan--Stirling equation of state, 
\begin{equation}
    \frac{1}{N} \beta F_\mathrm{hs} = \ln(\rho \Lambda^3) - 1 + \frac{\eta (4 - 3 \eta)}{(1 - \eta)^2}, 
\label{eq:F_hs}
\end{equation}
where $\eta=\picst\rho d^3/6$ is the packing fraction of hard-sphere system and $\Lambda$ is the de Broglie thermal wavelength \cite{hansen_theory_2013}. The effective hard-sphere diameter, $d$, is determined by Eq.~(\ref{eq:d_hs}). 

Its second term, which involves the perturbation potential, can be treated by the Laplace transform techniques proposed by Lutsko and Nicolis \cite{lutsko_effect_2005}. Below is a step-by-step guide for calculating this integral.

We start by decomposing the integral as
\begin{equation}
    \int_0^\infty r^2g_d(r)w(r)\,\mathrm{d}r = \int_0^\infty r^2g_d(r)v(r)\,\mathrm{d}r - \int_0^\infty r^2g_d(r)v_0(r)\,\mathrm{d}r .
\label{eq:int_gw}
\end{equation}
The integrand of the second term, which involves the potential $v_0(r)$, is non-zero only over a very short range $[d,r_0]$. We employ the model of Verlet and Weis \cite{verlet_equilibrium_1972} for the hard-sphere pair distribution function: 
\begin{widetext}
\begin{equation}
        g_d(r) = \left\{
    \begin{aligned}
        & 0, \quad r \leq d   \\[7pt]
        & g^\mathrm{PY}_{d_0}(r) + \frac{C}{r}\exp{\big(-m(r-d)\big)}\cos{\big(-m(r-d)\big)}, \quad r>d  
    \end{aligned}
    \right. 
    .
\end{equation}
\end{widetext}
Here, $g^\mathrm{PY}_{d_0}(r)$ is the Percus--Yevick (PY) pair distribution function for the hard-sphere system with a diameter $d_0=d(1-\eta/16)^{1/3}$. The parameters $C$ and $m$ are determined using the following relations  \cite{verlet_equilibrium_1972,henderson_direct_1975}: 
\begin{equation}
    \frac{C}{d}=\frac{2-\eta}{2(1-\eta)^3} - g^\mathrm{PY}_{d_0}(d)
\end{equation}
and
\begin{equation}
    md \approx \frac{24\,C/d}{\eta_0 \; g^\mathrm{PY}_{d_0}(d)} ,
\end{equation}
where $\eta_0=\picst\rho d_0^3/6$. Near $r \approx d_0$, $g^\mathrm{PY}_{d_0}(r)$ can be approximated by \cite{henderson_direct_1975}
\begin{equation}
    g^\mathrm{PY}_{d_0}(r) = \frac{2+\eta_0}{2(1-\eta_0)^2} - \frac{9}{2}\eta_0\frac{1+\eta_0}{(1-\eta_0)^3}(\frac{r}{d_0}-1) .
\label{eq:gPYd0}
\end{equation}
Since the interval $[d,r_0]$ is close to $d_0$, the second term in Eq.~(\ref{eq:int_gw}) can be calculated readily.

Next, we rewrite the first term in Eq.~(\ref{eq:int_gw}), 
\begin{widetext}
\begin{equation}
    \begin{aligned}
         \int_0^\infty r^2g_d(r)v(r)\,\mathrm{d}r = 
             \int_0^\infty r^2 g^\mathrm{PY}_{d_0}(r) v(r)\,\mathrm{d}r   
             - \int_{d_0}^d r^2 g^\mathrm{PY}_{d_0}(r) v(r)\,\mathrm{d}r  
             + \int_d^\infty r\,C\exp{\big(-m(r-d)\big)}\cos{\big(-m(r-d)\big)} v(r)\,\mathrm{d}r .
    \end{aligned}
\end{equation}
\end{widetext}
Its second integral can be calculated with the help of Eq.~(\ref{eq:gPYd0}). Its third integral can be easily evaluated numerically, or even analytically for $v(r)$ in some proper forms. Its first integral can be computed using the inverse Laplace transform of $rv(r)$,
\begin{equation}
    rv(r) = \mathscr{L}\{V(s)\}=\int_0^\infty V(s) \exp{(-sr)} \, \mathrm{d}s ,
\end{equation}
so 
\begin{eqnarray}
    \int_0^\infty r^2 g^\mathrm{PY}_{d_0}(r) v(r)\,\mathrm{d}r && = \int_0^\infty r\, g^\mathrm{PY}_{d_0}(r) \int_0^\infty V(s) \exp{(-sr)} \, \mathrm{d}s\,\mathrm{d}r \nonumber    \\[7pt]
    && = \int_0^\infty V(s) \int_0^\infty r\, g^\mathrm{PY}_{d_0}(r) \exp{(-sr)} \, \mathrm{d}r\,\mathrm{d}s \nonumber    \\[7pt]
    && = \int_0^\infty V(s)G_{d_0}^\mathrm{PY}(s) \,\mathrm{d}s , 
\end{eqnarray}
where $G_{d_0}^\mathrm{PY}(s)$ is the Laplace transform of $r g^\mathrm{PY}_{d_0}(r)$, given by Wertheim \cite{wertheim_exact_1963}: 
\begin{equation}
    G_{d_0}^\mathrm{PY}(s) =  \frac{d_0^2 \cdot (sd_0) \cdot L_{d_0}(sd_0)}{12\eta_0 \big(L_{d_0}(sd_0) + S_{d_0}(sd_0)\exp{(sd_0)}\big)}, 
\end{equation}
with $L_{d_0}(x) = 12\eta_0\big((1+\frac{1}{2}\eta_0)x + 1 + 2\eta_0\big)$ and $S_{d_0}(x) = (1-\eta_0)^2 x^3 + 6\eta_0(1-\eta_0) x^2 + 18\eta_0^2 x - 12\eta_0(1+2\eta_0)$. 

Finally, to compute the Helmholtz free energy, we need $V(s)$, which may have an analytical form for some simple potentials. For Lennard-Jones potential, $V(s)$ is given by 
\begin{equation}
    V_\mathrm{LJ}(s) = \epsilon\sigma^2\Big(-\frac{(s \sigma)^4}{6} + \frac{(s \sigma)^{10}}{907200}\Big) .
\end{equation}

If no analytical form of $V(s)$ is available, numerical methods such as numerical inverse Laplace transform can be used to evaluate it.

\section{Dimensionless form of nucleation rate}
\label{Appendix:Dimensionless}
\normalsize The steady-state nucleation rate, in units of per volume per time, is expressed as 
\begin{eqnarray}
    I =&& \dfrac{\rho_0}{\displaystyle\int_\mathrm{subC} \exp{\big(-\frac{\Delta\varOmega(\boldsymbol{x})}{k_\mathrm{B}T}\big) q(\boldsymbol{x}) \,\mathrm{d}\boldsymbol{x}}} \times \exp{\big(-\dfrac{\Delta\varOmega_\mathrm{c}}{k_\mathrm{B}T}\big)} \nonumber \\
    &&\times \big(2\picst k_\mathrm{B}T\big)^{\frac{n-2}{2}} \times\frac{|{\lambda_\mathbf{Z}}_1|}{\sqrt{|\det \mathbf{H}|}} .
\label{eq:I(Appendix)}
\end{eqnarray}
To derive a dimensionless version, we select appropriate scaling parameters based on the potential under study. For Lennard-Jones-like potentials, the following reductions parameters are used: energy $\epsilon$, length $\sigma$, monomer mass $m$, and the Boltzmann constant $k_\mathrm{B}$. The time reduction scale is derived from these parameters: $\tau = \sqrt{m\sigma^2/\epsilon}$.
Consequently, we define the reduced radius $R^*=R/\sigma$, the reduced density $\rho^*=\rho\sigma^3$, the reduced time $t^*=t/\tau$, the reduced temperature $T^*=Tk_\mathrm{B}/\epsilon$, and the reduced work of formation $\Delta\varOmega^* = \Delta\varOmega/\epsilon$.
It follows directly that
\begin{equation*}
    \exp{\big(-\dfrac{\Delta\varOmega_\mathrm{c}}{k_\mathrm{B}T}\big)} = 
    \exp{\big(-\dfrac{\Delta\varOmega^*_\mathrm{c}\cdot\epsilon}{k_\mathrm{B}T^*\cdot\epsilon/k_\mathrm{B}}\big)} = 
    \exp{\big(-\dfrac{\Delta\varOmega^*_\mathrm{c}}{T^*}\big)}
\end{equation*}
and
\begin{equation*}
    \big(2\picst k_\mathrm{B}T\big)^{\frac{n-2}{2}} = 
    \big(2\picst k_\mathrm{B}T^*\cdot\epsilon/k_\mathrm{B}\big)^{\frac{n-2}{2}} = 
    \big(2\picst T^*\big)^{\frac{n-2}{2}} \cdot \epsilon^{\frac{n-2}{2}} .
\end{equation*}

Using $[x]$ to represent the reduction parameter of a phase space element and defining $\big[\mathbf{X}\big] = \text{diag}([x_1], \dots , [x_n])$, we have $\boldsymbol{x} = \big[\mathbf{X}\big]x^*$. We can rewrite Eq.~(\ref{eq:dotxDH_Alekseechkin}) as
\begin{equation*}
    \pypx{(\big[\mathbf{X}\big]\boldsymbol{x^*})}{(t^*\cdot\tau)} = -\frac{1}{k_\mathrm{B}T^*\cdot\epsilon/k_\mathrm{B}}\mathbf{Z}\big[\mathbf{X}\big](\boldsymbol{x^*}-\boldsymbol{x^*}_\mathrm{c}) .
\end{equation*}
To match the reduced form
\begin{equation*}
        \pypx{\boldsymbol{x^*}}{t^*} =  -\frac{1}{T^*}\mathbf{Z^*}(\boldsymbol{x^*}-\boldsymbol{x^*}_\mathrm{c}) ,
\end{equation*}
we find that $\mathbf{Z^*} =\big[\mathbf{X}\big]^{-1}\mathbf{Z}\big[\mathbf{X}\big]\cdot\tau/\epsilon$ and ${\lambda_\mathbf{Z}}_1^* = {\lambda_\mathbf{Z}}_1\cdot\tau/\epsilon$.

The Hessian matrix and its reduced form are related by
\begin{equation*}
    \mathbf{H} = \epsilon \big[\mathbf{X}\big]^{-1} \mathbf{H^*} \big[\mathbf{X}\big]^{-1}.
\end{equation*}
Therefore,  
\begin{equation*}
    \det \mathbf{H} = \det \mathbf{H^*}\cdot\frac{\epsilon^n}{\Big(\prod \limits_{i=1}^n[x_i]\Big)^2} .
\end{equation*}

The first factor in Eq.~(\ref{eq:I(Appendix)}), which is indeed the constant $f_0$ from Eq.~(\ref{eq:f0}), can be rewritten as
\begin{widetext}
\begin{equation*}
    \dfrac{\rho_0}{\displaystyle\int_\mathrm{subC} \exp{\big(-\frac{\Delta\varOmega(\boldsymbol{x})}{k_\mathrm{B}T}\big) q(\boldsymbol{x}) \,\mathrm{d}\boldsymbol{x}}} = 
    \dfrac{\rho_0^*}{\displaystyle\int_\mathrm{subC} \exp{\big(-\frac{\Delta\varOmega^*(\boldsymbol{x^*})}{T^*}\big) q(\boldsymbol{x^*}) \,\mathrm{d}\boldsymbol{x^*}}} \cdot\frac{1}{\sigma^3 \prod \limits_{i=1}^n[x_i]} .
\end{equation*}
\end{widetext}
It should also be mentioned that Alekseechkin's formulation for $f_0$, i.e., Eq.~(\ref{eq:f0_Alekseechkin}), yields the same reduced form, $f_0 = f_0^* /(\sigma^3 \prod \limits_{i=1}^n[x_i])$.

Thus, the reduced nucleation rate becomes 
\begin{eqnarray}
    I^* =&& \dfrac{\rho_0^*}{\displaystyle\int_\mathrm{subC} \exp{\big(-\frac{\Delta\varOmega^*(\boldsymbol{x^*})}{T^*}\big) q(\boldsymbol{x^*}) \,\mathrm{d}\boldsymbol{x^*}}} \times \exp{\big(-\dfrac{\Delta\varOmega^*_\mathrm{c}}{T^*}\big)} \nonumber   \\
    &&\times \big(2\picst T^*\big)^{\frac{n-2}{2}} \times \frac{|{\lambda_\mathbf{Z}}_1^*|}{\sqrt{|\det \mathbf{H^*}|}} ,
\label{eq:I_LJreduced}
\end{eqnarray}
where $I^* = I\sigma^3\tau$.

An analogous derivation applies to the DDFT equations. The essential results are 
\begin{equation*}
\begin{aligned}
    &\mathbf{g} = \sigma^2 \big[\mathbf{X}\big]^{-1} \mathbf{g^*} \big[\mathbf{X}\big]^{-1}, \\[11pt]
    &D_\mathrm{diff} = D_\mathrm{diff}^* \cdot \frac{\sigma^2}{\tau} ,   \\[11pt]
    &\mathbf{g^*}\pypx{\boldsymbol{x^*}}{t^*} = -\dfrac{D_\mathrm{diff}^*}{T^*}\pypx{\Delta\varOmega^*}{\boldsymbol{x^*}} = -\dfrac{D_\mathrm{diff}^*}{T^*}\boldsymbol{\nabla^*}\Delta\varOmega^* .
\end{aligned} 
\end{equation*}

\section{Nucleation driving force in classical nucleation theory}
\label{Appendix:Delta mu in CNT}
In the framework of CNT, the nucleation driving force is usually expressed as the chemical potential difference between two phases, denoted as $\Delta\mu$. It is important to note its definition: the difference between the vapor chemical potential $\mu^\mathrm{v}$ and the liquid chemical potential $\mu^\mathrm{l}$ \emph{at the vapor pressure} $p^\mathrm{v}$ and temperature $T$ \cite{kalikmanov_nucleation_2013}, i.e., 
\begin{equation}
    \Delta\mu = \mu^\mathrm{v}(p^\mathrm{v},T) - \mu^\mathrm{l}(p^\mathrm{v},T). 
\end{equation}
While $\mu^\mathrm{v}(p^\mathrm{v},T)$ can be directly determined, obtaining $\mu^\mathrm{l}(p^\mathrm{v},T)$ requires the Gibbs-Duhem equation: 
\begin{equation*}
    s\mathrm{d}T-v\mathrm{d}p+\mathrm{d}\mu=0,
\end{equation*}
where $s$ is the entropy per particle, $v$ is the volume per particle, and $p$ is the pressure. Since the process is assumed to be isothermal, the temperature dependence can be omitted, yielding  
\begin{equation*}
    \mu^\mathrm{l}(p^\mathrm{v}) = \mu^\mathrm{l}(p^\mathrm{l}) + \int_{p^\mathrm{l}}^{p^\mathrm{v}}\frac{1}{\rho}\mathrm{d}p .
\end{equation*}
Thus, the expression for $\Delta\mu$ becomes 
\begin{equation}
    \Delta\mu = \mu^\mathrm{v}(p^\mathrm{v}) - \mu^\mathrm{l}(p^\mathrm{l}) - \int_{p^\mathrm{l}}^{p^\mathrm{v}}\frac{1}{\rho}\mathrm{d}p .
\label{eq:Delta mu}
\end{equation}

In this work, we define the driving force based on the $\mathcal{F}\text{-}\rho$ curve, where the slope corresponds to the chemical potential $\mu$. CNT assumes that nucleation occurs under the condition of maximal driving force,  which is achieved when $\mu(\rho_0)=\mu(\rho^\mathrm{l})$. This corresponds to the case where $\mu^\mathrm{v}(p^\mathrm{v}) = \mu^\mathrm{l}(p^\mathrm{l})$ in Eq.~(\ref{eq:Delta mu}). Furthermore, assuming the liquid is incompressible, we obtain 
\begin{equation}
    \Delta\mu = \frac{1}{\rho^\mathrm{l}}(p^\mathrm{l} - p^\mathrm{v}) .
\end{equation}
Using thermodynamic relations, we express the pressure as 
\begin{equation}
    p = \rho^2\big(\frac{\partial (\mathcal{F}/\rho)}{\partial \rho}\big) = \mu\rho - \mathcal{F}, 
\end{equation}
which is indeed the negative of the grand potential density. Consequently, $p^\mathrm{l} - p^\mathrm{v} = g_n(\rho^\mathrm{l})$. Finally, assuming $\rho^\mathrm{l} \approx \rho_2^\mathrm{eq}$, we arrive at $\Delta\mu = g_n^\mathrm{CNT}/\rho_2^\mathrm{eq}$.

\section{Invariance of the negative eigenvalue of matrix $\mathbf{Z}$}
\label{Appendix:lambdaZ}
Matrix $\mathbf{Z}$ has exactly one negative eigenvalue, ${\lambda_\mathbf{Z}}_1$, with its corresponding eigenvector, ${\boldsymbol{v}_\mathbf{Z}}_1$, indicating the nucleation flux direction, i.e., the unstable direction in the phase space. A key property of this eigenvalue is its invariance: ${\lambda_\mathbf{Z}}_1$ remains unchanged under any reparameterization along the same unstable direction. This invariance can be demonstrated as follows.

Suppose $x'_1$ and $y'_1$ are two distinct parameterizations along the unstable direction. Their time evolution near the saddle point can be expressed as $\frac{\partial x'_1}{\partial t} = -\frac{{\lambda^x_\mathbf{Z}}_1}{k_\mathrm{B}T}(x'_1 - {x'_1}_\mathrm{c})$ and $\frac{\partial y'_1}{\partial t} = -\frac{{\lambda^y_\mathbf{Z}}_1}{k_\mathrm{B}T}(y'_1 - {y'_1}_\mathrm{c})$, where ${x'_1}_\mathrm{c}$ and ${y'_1}_\mathrm{c}$ are the critical values corresponding to the saddle point under the two parameterizations. Treating $y'_1$ as a function of $x'_1$ and applying the chain rule gives
\begin{equation}
    \frac{\partial y'_1}{\partial t} = \frac{\partial y'_1}{\partial x'_1} \frac{\partial x'_1}{\partial t} = -{\lambda^x_\mathbf{Z}}_1 \cdot \frac{\partial y'_1}{\partial x'_1} \cdot (x'_1 - {x'_1}_\mathrm{c}).
\label{eq:Invariance_lambdaZ1}
\end{equation}
Near the saddle point, we can approximate the relationship between the variables via a linear expansion: 
\begin{equation*}
    (y'_1 - {y'_1}_\mathrm{c}) \approx \frac{\partial y'_1}{\partial x'_1} \cdot (x'_1 - {x'_1}_\mathrm{c}).
\end{equation*}
Substituting this into Eq.~(\ref{eq:Invariance_lambdaZ1}) yields
\begin{equation*}
    \frac{\partial y'_1}{\partial t} = -\frac{{\lambda^x_\mathbf{Z}}_1}{k_\mathrm{B}T}(y'_1 - {y'_1}_\mathrm{c}),
\end{equation*}
which confirms that ${\lambda^y_\mathbf{Z}}_1 = {\lambda^x_\mathbf{Z}}_1$. Hence, the eigenvalue associated with the same unstable direction is invariant under changes of parameterization. Indeed, $|{\lambda_\mathbf{Z}}_1| / (k_\mathrm{B}T)$ is the exponential growth rate of the system along the unstable direction. 

A direct application of this result is found in the one-dimensional CNT. Whether one uses the cluster radius or the number of monomers as the variable, the value of ${\lambda_\mathbf{Z}}_1^\mathrm{CNT}$ remains the same as given by Eq.~(\ref{eq:lambda_CNT}). This invariance can also be verified by explicitly calculating the components of ${\lambda_\mathbf{Z}}_1^\mathrm{CNT}$ in both variable choices. 

An important implication for the multivariable nucleation models is the following: if two models yield the same unstable direction in their phase spaces, they must have the same value of ${\lambda_\mathbf{Z}}_1$. In particular, as the number of physically meaningful variables increases, the value of ${\lambda_\mathbf{Z}}_1$ in a multivariable model under SGA should asymptotically approach that of the SGA diffuse-interface model, because the SGA diffuse-interface model provides the most precise description of the density profile. 

\bibliography{References}

\end{document}